\documentclass[journal,comsoc,10pt]{IEEEtran}
\usepackage{cite,graphicx,amssymb,amsmath,color,textcomp,tabularx}
\usepackage{multicol}
\usepackage[cmintegrals]{newtxmath}
\usepackage{graphicx}
\usepackage{stfloats}
\usepackage{algorithm,algorithmic}
\usepackage{multirow}
\usepackage{xcolor}
\usepackage{lettrine}

\IEEEoverridecommandlockouts

\begin{document}
\title{Non-orthogonal Multiple Access Assisted Multi-Region Geocast}
\author{{Yi~Zhang,~\IEEEmembership{Student Member,~IEEE}, Hui-Ming~Wang,~\IEEEmembership{Senior Member,~IEEE}, Zhiguo~Ding,~\IEEEmembership{Senior Member,~IEEE}, and Moon Ho Lee,~\IEEEmembership{Life Senior Member,~IEEE}}
\thanks{Yi Zhang and Hui-Ming Wang are with the School of Electronic and Information Engineering, Xi'an Jiaotong University, Xi'an, 710049, Shaanxi, P. R. China (e-mail: yi.zhang.cn@outlook.com; xjbswhm@gmail.com). Yi Zhang is with the Department of Electrical and Computer Engineering, University of Texas at Austin, TX, US. Zhiguo Ding is with the School of Computing and Communications, Lancaster University, LA1 4YW, UK (e-mail: z.ding@lancaster.ac.uk). Moon Ho Lee is with the Department of Electronics Engineering, Chonbuk National University, Jeonju 561-756, South Korea (e-mail: moonho@jbnu.ac.kr). The corresponding author is Hui-Ming Wang.}
\thanks{The work was supported by the National Natural Science Foundation of China under Grants 61671364, the Author of National Excellent Doctoral Dissertation of China under Grant 201340,
and the Young Talent Support Fund of Science and Technology of Shaanxi Province under Grant 2015KJXX-01. The work of Z. Ding was supported by the UK EPSRC under grant number EP/N005597/1 and by H2020-MSCA-RISE-2015 under grant number 690750.}
\thanks{This paper has been accepted in part for presentation at IEEE Wireless Communications and Networking Conference (WCNC 2017), San Francisco, USA \cite{Conf_Version}.}}
\maketitle

\begin{abstract}
This work proposes a novel \textit{location-based} multi-group multicast framework which is termed as \textit{non-orthogonal multiple access (NOMA) assisted multi-region geocast}. This novel spectrum sharing framework exploits the NOMA technology to realize the simultaneous delivery of different messages to different user groups characterized by different \emph{geographical locations}. The essence of the proposed framework is that the geographical information of user groups unites NOMA and multi-group multicast to enhance the spectral efficiency (SE) and the energy efficiency (EE) of wireless transmissions. Specifically, we investigate the downlink beamforming design of the proposed framework in multiple-input single-output (MISO) settings. The decoding strategy for NOMA is designed and guaranteed by users' geographical information and required quality of service (QoS). The majorization and minimization (MM) algorithm is exploited to solve the non-convex and intractable problems therein. Comprehensive numerical experiments are further provided to show that NOMA holds tremendous promise but also limitations in terms of SE and EE, compared with spatial division multiple access (SDMA) and orthogonal multiple access (OMA).
\end{abstract}

\begin{IEEEkeywords}
Non-orthogonal multiple access, multi-group multicast, geographical information, design of decoding order, quality of service.
\end{IEEEkeywords}

%INTRODUCTION AND MOTIVATION
\section{Introduction}\label{Sec_Intro}
%Introduction of Motivation
\IEEEPARstart{W}{ireless} spectrum has become severely limited and the 5G system is expected to offer 10 times spectral efficiency (SE) and energy efficiency (EE) in comparison with the 4G system\cite{5G_SE_EE}. The anticipated thousand-fold increase in wireless data traffic urgently calls for highly efficient spectrum utilization to achieve these challenging goals. Various new techniques have been proposed, among which multi-group multicast and non-orthogonal multiple access (NOMA) are two promising ones.

%Multi-group multicast
Wireless multi-group multicast is considered as an effective solution to improve SE and EE by exploiting the idiosyncracies of wireless medium\cite{Luo_Multicast}. It utilizes the available bandwidth for delivering the same content to multiple users and further exploits the spatial diversity by simultaneously serving multiple user groups, which significantly promotes the efficient utilization of the spectrum\cite{Multicast}. Besides, transmit beamforming makes it possible to focus energy into the directions of the target users within a specified geographical region\cite{Luo_MultiGroupMulticast}.

%Noma
NOMA is widely recognized as another emerging paradigm to enhance spectrum utilization for the 5G system\cite{NOMA_Dai}. The major advantage of NOMA is to encourage spectrum sharing among users in a non-orthogonal way. Different from conventional orthogonal multiple access (OMA), NOMA realizes the simultaneous transmission of multiple data-streams via power domain division \cite{Noma_Performace}. At a transmitter, a superposition of different messages is broadcasted. At receivers, successive interference cancellation (SIC) is used to realize multi-user separation and detection.

%Common Advantage
Current designs from industry and academia consider multi-group multicast and NOMA independently one from the other and for different intentions. However, there is an important bridge between them: \emph{geographical locations} of the target users plays a critical role in both the two techniques. For multi-group multicast, user grouping is quite important in satisfying concurrent users' requests for the same content. Thus, it is natural to exploit multi-group multicast for transmitting different messages to user clusters in different geographical regions that could be intuitively characterized by the \emph{distances} from the transmitter to the users. On the other hand, the nature of NOMA is to utilize the signal-to-interference-plus-noise ratio (SINR) imparity among users that results from either the natural near-far effect or the non-uniform power allocation at a transmitter \cite{NOMA_Dai,Noma_Performace}. Accordingly, the difference in channel gains, caused by different signal attenuation due to different \textit{distances} of users to the transmitter, is a natural advantage for NOMA. We can see that physical  \textit{\textbf{distance}} becomes a bridge that connects multi-group multicast and NOMA, because distance is not only an important geographical feature of regions, but also a crucial factor which makes the users in different regions have different channel gains.

%Our Proposed NOMA Assisted Multi-Region Geocast
Clearly, it is of paramount importance to use the geographical information for designing multi-group multicast and NOMA altogether. Out of this we porpose a novel multi-group multicast framework in this paper which is termed as \textit{NOMA assisted multi-region geocast}\footnote{It is worth to point out that the term \textit{geocast} derives from a technique for routing protocols which realizes the delivery of a common message to a group of users within a specified geographical region\cite{Geocast3}.}. The novelty and particularity of this new framework consists in the exploitation of geographical features of the target users in different geographical regions and broadcast properties of wireless medium. To be specific, our proposed NOMA assisted multi-region geocast refers to the simultaneous transmission of different messages to user clusters located in different predetermined regions via NOMA principles, and the users in the same region are interested in a common message. By exploiting multiple antenna technology, the transmit beamforming, aided by the geographical information of users, can provide a new manner to fully utilize the limited spectrum and to serve users with the minimum possible energy consumption.

%Motivation
With NOMA assisted multi-region geocast, many new services and applications are feasible. These include geographical command systems, emergency communications for public alerting, and commercial geographical messaging applications, such as position-based advertising, weather forecasting, and traffic advisory services. These imply that our proposed NOMA assisted multi-region geocast is a promising \emph{location-based} spectrum sharing framework, which holds tremendous potential to support the increasing demand for higher quality services with the minimum possible energy expenditure for the future 5G system.

%Related Works
\subsection{Related Works and Motivations}
%Related Works With Multicast
The transmit beamforming for multi-group multicast have been initially solved with the semi-definite relaxation (SDR) method in \cite{Luo_Multicast,Luo_MultiGroupMulticast}. Further, a weighted max-min fair multi-group multicast problem has been studied in \cite{FairMulticast} under per-antenna power constraints. In \cite{MultigroupMulticast}, the distributed beamforming design has been attempted for a multi-group multicasting relay network. However, all these works have neither exploited users' geographical information nor took the advantage of NOMA to further improve the SE and EE of wireless transmissions.

%Related Works on NOMA Plus Decoding Order Problems
Current works on NOMA primarily focused on the SE improvement in either single-input single-output (SISO) systems\cite{Fairness_NOMA, CTS, CRS, PLS_NOMA} or multiple-input multiple-output (MIMO) systems that can be decoupled into multiple SISO NOMA sub-systems by user clustering \cite{MIMO_NOMA1} or signal alignment \cite{MIMO_NOMA2}. Only some works have directly tackled NOMA in multiple-antenna settings due to the difficulty of designing decoding order for SIC at receivers. In \cite{NOMA_Jchoi}, multicast design has been investigated in a multiple-input single-output (MISO) NOMA system wherein a transmitter broadcasts \textit{only two} different data streams. In \cite{JointBPowerAllo}, the beamforming design has been considered in MISO NOMA system where users are partitioned into multiple cluster with uniform power allocation. In \cite{MISOSubcarrier}, the optimal resource allocation for multicarrier downlink MISO NOMA system has been investigated. In \cite{RobustBeamform}, the worst-case achievable sum rate has been considered and an alternative optimization algorithm was proposed. In \cite{hybrid}, a Hybrid NOMA precoding algorithm has been proposed by applying the quasi-degradation into downlink MISO NOMA system. Under bounded channel uncertainties, a robust beamforming techniques has been provided for MISO NOMA system in \cite{RobustBeamform2}. Besides, in \cite{MISO_NOMA}, the sum rate maximization problem has also been studied in a downlink MISO NOMA system, in which the power allocation policy is to allocate more power to users with worse channel conditions. However, this widely used policy is unnecessary and even degrades the overall performance of NOMA. Accordingly, in this work, we directly deal with NOMA in MISO systems and propose to guarantee users' quality of service (QoS) with minimum SINR thresholds. Our previous work \cite{Conf_Version} has framed this NOMA assisted multi-region geocast by just focusing on its SE perspective. 

In addition, NOMA can be viewed as a special form of cognitive radio (CR) networks\cite{CognitiveRadios2}, in which the user having stronger channel gains is viewed as a secondary user while the user having poorer channel gains is viewed as a primary one. In \cite{CognitiveNOMA}, the above described CR inspired NOMA framework was generalized by exploiting multiple antenna technologies and QoS guarantees for weaker users.

Currently, only some recent work have studied the EE issues in SISO and MIMO NOMA systems with statistical CSI \cite{EE1, EE2, EE3}, which greatly motivates us to study NOMA from the perspectives of both SE and EE.
%No comparison with SDMA
Furthermore, current designs of NOMA were generally compared to the traditional OMA scheme\cite{MIMO_NOMA2,MISO_NOMA,EE1}, which obviously demonstrates the superiority and advantage of NOMA. However, conventional spatial division multiple access (SDMA) should be considered as a strong competitor when studying NOMA in multiple-antenna settings.

%Conclusion of motivation
Based on the above survey, this NOMA assisted multi-region geocast framework has a potential future for realizing higher efficient spectrum utilization. Further, comparisons between NOMA and SDMA are still not clear. Thereby, it is of great meaning to study this new framework from the perspectives of SE and EE for sake of further revealing the idiosyncrasy of NOMA.

%Contribution
\subsection{Our Work and Contributions}
In this work, with advanced beamforming techniques, we apply NOMA to simultaneously transmit superposed messages to multiple geographical regions, which germinates a location-based spectrum sharing framework. The main contributions and motivation are summarized as below:

\begin{enumerate}
	\item A novel multi-group multicast framework is proposed by exploiting both NOMA and the geographical information of users in transmit beamforming designs. Within the proposed framework, we initially minimize the total transmit power subject to an individual QoS constraint for each user. Secondly, we maximize the system sum rate under a predefined fair transmission policy that ensures constant ratios between the data rates of different messages. Finally, an EE maximization problem is investigated in the proposed framework under a dynamical access scenario. The majorization and minimization (MM) algorithms are employed to solve the \textit{non-convex} and \textit{intractable} problems therein.
	\item Different from previous works on NOMA, this work abolishes the conventional power allocation policy that more transmit power should be given to users with worse channel conditions. Instead, a new decoding order for SIC is designed and guaranteed by users' geographical information and their required QoS (SINR threshold) in the process of transmit beamforming designs.
	\item Comprehensive numerical experiments are provided, with respect to different key parameters, to show that NOMA scheme holds {\textit{tremendous promise}} for enhancing SE and EE but also {\textit{certain limitations}} in wireless transmissions, compared with both conventional SDMA and OMA schemes. Our findings aim to provide not only insights on possibilities of NOMA but also guidance on applications of NOMA in future 5G systems.
\end{enumerate}

%Organization and Notations
\subsection{Organization and Notations}
%Organization
\emph{Organization}: In Section \ref{SecModel}, we present the system and the signal model of multi-region geocast in a MISO NOMA system. In Section \ref{SecQoS}, we minimize the total transmit power of the system subject to an individual QoS constraint for each user. In Section \ref{SecFair}, we further maximize the sum rate of the system under a predefined fair transmission policy that ensures constant ratios between the data rates of different messages. In Section \ref{SecEE}, we lastly investigate an EE optimization problem in the previous system with an upgrade that an extra user group having better channel condition is simultaneously served. Section \ref{SecSim} provides simulation results. Concluding remarks are made in Section \ref{SecCon}.

%Notations
\emph{Notations}: Boldface uppercase letters and boldface lowercase letters are used to denote matrices and column vectors, respectively. $(\cdot)^T$, $(\cdot)^*$, $(\cdot)^H$, $\textrm{Tr}(\cdot)$ and $\mathbb{E}(\cdot)$ are the transpose, conjugate, Hermitian transpose, trace, and expectation operators. x $\sim\mathcal{CN}(\mathbf{0},\mathbf{I}_M)$ indicates that x is a circular symmetric complex Gaussian random vector whose mean vector is $\mathbf{0}$ and covariance matrix is $\mathbf{I}_M$. $\parallel\cdot\parallel_2$ represents the Euclidean norm of a vector. $\textrm{Re}\{\cdot\}$ and $\textrm{Im}\{\cdot\}$ is the real part and the imaginary part of a complex number, respectively.
$\textrm{rank}(\cdot)$ represents the number of linearly independent rows or columns of a matrix.
$\mathcal{O}(\cdot)$ denotes the complexity.

\section{System and Signal Models}\label{SecModel}
\subsection{System Model of NOMA Assisted Multi-Region Geocast}
%System Compositions
Consider a wireless transmission scenario which consists of a transmitter equipped with $M$ antennas and $G$ groups of users each equipped with a single antenna. Let $U_g$ denote the number of users in the $g$-th group for $1\leq g\leq G$. Note that the lowercase letter $g$ is used to represent the group index, and the lowercase letter $u$ is used to represent the user index in an arbitrary group, where $1\leq u\leq U_g$ and $1\leq g\leq G$.

%Explanation of how to distinguish the different groups (regions)
\begin{figure}[!t]
    \centering
        \includegraphics[scale = 0.18]{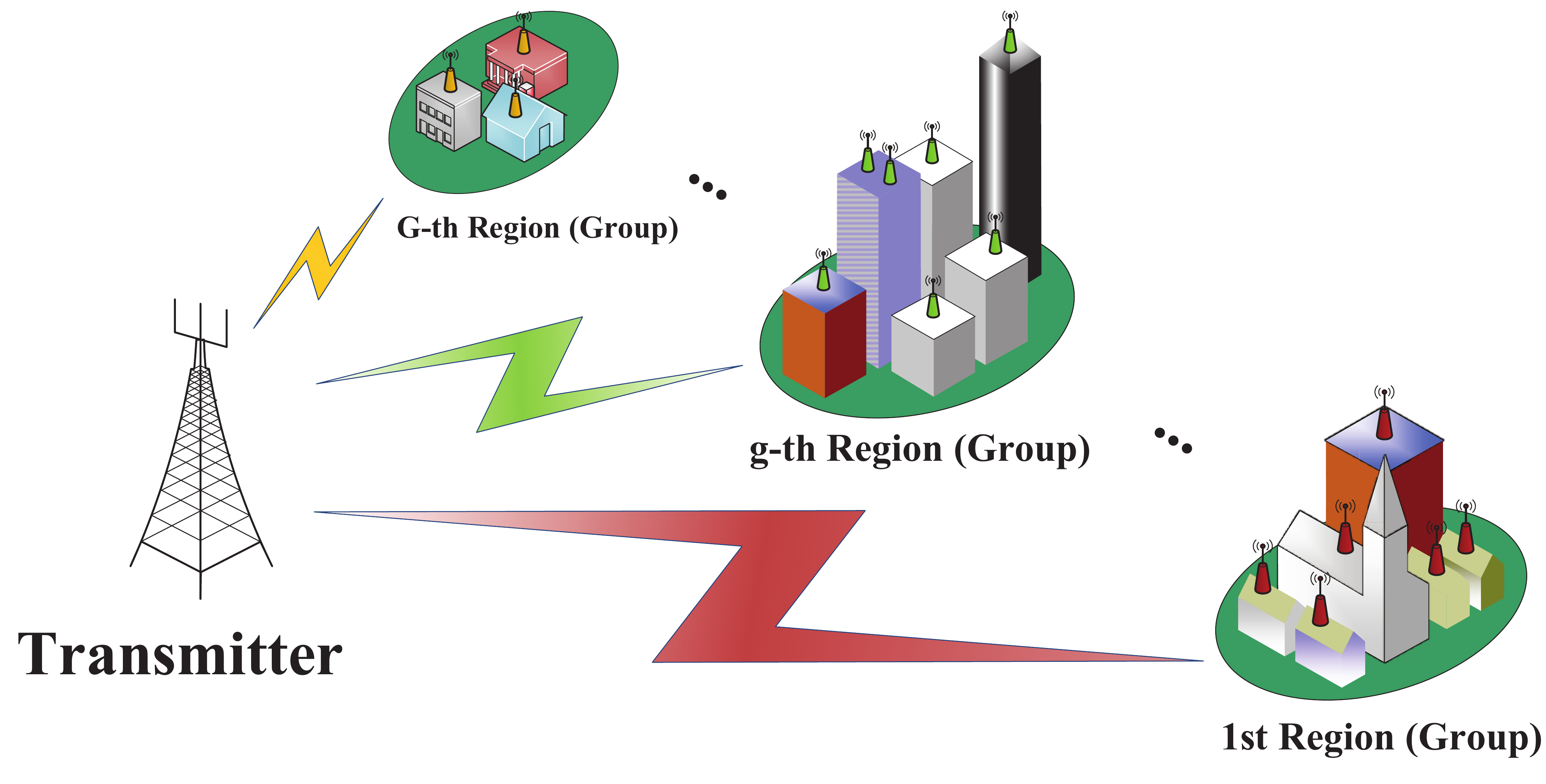}
    \caption{System Model of NOMA Assisted Multi-Region Geocast}
    \label{fig:SystemModel}
\end{figure}

As shown in Fig. \ref{fig:SystemModel}, each group is located in a predetermined region different from the other groups. Therefore, there is a one-to-one correspondence between $G$ groups and $G$ regions. It is worth to point out that these $G$ regions are distinct from each other by their average distances to the transmitter, which is explained as follows.
We use $d_g^\textrm{min}$ to denote the minimum distance from the transmitter to the $g$-th region and $d_g^\textrm{max}$ to denote the maximum distance. Without regard to the specific shape of the region, we suppose that the $U_g$ users are randomly deployed in the $g$-th region. Then, we can define $d_g^\textrm{ave} = \frac{d_g^\textrm{min}+d_g^\textrm{max}}{2}$ as the average distance between the transmitter and the $g$-th region. Without loss of generality, assume that $\left\{d_g^\textrm{ave}\right\}_{g=1}^G$ are sorted as $d_1^\textrm{ave}\geq d_{2}^\textrm{ave}\geq...d_{G-1}^\textrm{ave}\geq d_G^\textrm{ave}\geq0$, which indicates that the $G$-th region is the nearest one to the transmitter and the 1st region is the farthest one.

%Channel Model
The channel between the transmitter and each user is supposed to be a composition of large-scale path loss and quasi-static small-scale fading. Let $\mathbf{h}_{u,g}$ represent the $M\times1$ channel vector from the transmitter to the $u$-th user in the $g$-th group and be modeled as
\begin{equation}\label{channel_gain}
    \mathbf{h}_{u,g} = \mathbf{g}_{u,g}d_{u,g}^{-\frac{\alpha}{2}},~1\leq u\leq U_g,~1\leq g\leq G,
\end{equation}
where ${\mathbf{g}_{u,g}}\sim \mathcal{CN}(0,\mathbf{I}_M)$ is the small-scale fading coefficient, $d_{u,g}$ is the distance from the transmitter to the $u$-th user in the $g$-th group, and $\alpha$ is the path loss exponent. All users' instantaneous CSI is supposed to be available at the transmitter.

\subsection{Signal Model of NOMA Assisted Multi-Region Geocast}\label{SMNOMA}
%Signal Model
We let $s_g$ denote the geocast information symbol intended for all the $U_g$ users in the $g$-th group with $\mathbb{E}\left(\big|s_g\big|^2\right)=1$, and $\mathbf{w}_g$ represent the corresponding $M\times1$ transmit beamformer, where $1\leq g\leq G$. The transmitter broadcasts the summation of the $G$ weighted messages given by $\sum_{g=1}^G \mathbf{w}_gs_g$.
The users in the same group (or region) are interested in a common message while different groups desire different messages. Thereby, the received signal at the $u$-th user in the $g$-th group can be expressed as
\begin{equation}\label{Received Signal}
\begin{aligned}
&y_{u,g} = \underbrace{\mathbf{h}_{u,g}^H \mathbf{w}_gs_g}_\textrm{Desired Message} + \underbrace{\textstyle\sum_{i\neq g}^G \mathbf{h}_{u,g}^H \mathbf{w}_is_i}_\textrm{Inter-group Interference} + \underbrace{z_{u,g}}_\textrm{Additive Noise},\\
&~~~~~~~~~~~~~~~~~~~~~~~~~~~~~~~~~1\leq u\leq U_g,~1\leq g\leq G,
\end{aligned}
\end{equation}
where $z_{u,g}$ is the additive Gaussian noise with zero mean and variance $\sigma^2$ for all users. The second term at the right-hand side (R.H.S.) of (\ref{Received Signal}) is the inter-group interference, namely the other groups' target messages.

%Reasoning why use NOMA
Unlike conventional multi-group multicast (i.e. SDMA) wherein the users decode their own desired messages by treating the messages for the other groups as noise \cite{Luo_MultiGroupMulticast}, NOMA assisted multi-region geocast exploits the geographical information of users to aid the beamforming design at the transmitter. This is because, a remarkable difference in the average distances of the regions to the transmitter is likely to make the channel gains vary significantly among the users located in different regions, which is a natural advantage for the implementation of NOMA.
Recall that in NOMA, users often apply SIC to eliminate partial co-channel interference\cite{NOMA_Dai}. To realize SIC, the transmitter should establish a decoding order for all the users. In this work, the decoding order is predefined in accordance with the aforesaid order of \textit{\textbf{the average distances}} of the regions to the transmitter. To be specific, the users in the $g$-th group firstly decode the message for the $i$-th group for $i<g$, and then remove this message from their own received mixtures, in the order $i=1,2...g-1$; the messages for the $i$-th group for $i>g$ will be treated as noise.
%DIscussion on NOMA Decoding Ordering
It is worth to point out that the predefined decoding order herein may not be optimal, but it facilitates both the implementation of SIC and the beamforming design. This work focuses on the beamforming design under a predefined decoding order rather than optimizing the decoding order. Furthermore, no recent literature has reported the optimal decoding order in MISO settings due to its great difficulty\cite{MISO_NOMA}.

%Abolishment of Old Power Allocation Policy
A widely used power allocation policy in NOMA is to assign more power to users with worse channels\cite{MISO_NOMA,NOMA_SISO_DING}.
However, this policy is unnecessary and unfavorable for the overall performance of NOMA, which motivates us to annul the policy and propose to guarantee users' QoS by minimum SINR thresholds. The specific processing is provided in the following.

%Explain how SIC is performed in this work
With the signal model provided by (\ref{Received Signal}) and the predefined decoding order of SIC, the decoding ability of any user can be characterized by its SINR. Let $\textrm{SINR}_{u,g}$ represent the SINR of the $u$-th user in the $g$-th group to decode its own desired message, i.e., $s_g$, then $\textrm{SINR}_{u,g}$ can be given by
\begin{equation}\label{SINR_Each_Group}
        \textrm{SINR}_{u,g}=\frac{\left|\mathbf{h}_{u,g}^H \mathbf{w}_g\right|^2}{\sum_{k=g+1}^G\left|\mathbf{h}_{u,g}^H \mathbf{w}_k\right|^2+\sigma^2},~1\leq g\leq G,~1\leq u\leq U_g,
\end{equation}
which holds if and only if all the $U_g$ users in the $g$-th group are able to successfully decode the information for the $i$-th group, $1\leq i\leq g-1$. The guarantee of (\ref{SINR_Each_Group}), namely the successful implementation of SIC given the predefined decoding order, is discussed in the sequel. We further denote $\textrm{SINR}_{u,g}^i$ as the SINR of the $u$-th user in the $g$-th group to decode the message for the $i$-th group, where $1\leq i\leq g-1$ (the users in the $g$-th group will treat the message for the $i$-th group as noise for $i>g$), and then $\textrm{SINR}_{u,g}^i$ can be given by
\begin{equation}\label{SINR_Among_Group}
\begin{split}
&\textrm{SINR}_{u,g}^i=\frac{\left|\mathbf{h}_{u,g}^H \mathbf{w}_i\right|^2}{\sum_{k=i+1}^G\left|\mathbf{h}_{u,g}^H \mathbf{w}_k\right|^2+\sigma^2},\\
&~~~~~~~~~~~2\leq g\leq G,~1\leq u\leq U_g,~1\leq i\leq g-1.
\end{split}
\end{equation}
According to (\ref{SINR_Among_Group}), $\left\{\textrm{SINR}_{u,g}^i\right\}_{i=1}^{g-1}$ represent the ability of the $u$-th user in the $g$-th group to decode the messages intended for the groups located farther from the transmitter compared with itself. Let $\Gamma_{g}$ for $1\leq g\leq G$ represent the minimum required SINR for successfully decoding $s_g$, then the successful implementation of SIC given the predefined decoding order is guaranteed if and only if the following constraints are satisfied:
\begin{equation}\label{ValidSIC}
    \textrm{SINR}_{u,g}^i\geq \Gamma_{i},~2\leq g\leq G,~1\leq u\leq U_g,~1\leq i\leq g-1.
\end{equation}
We herein conclude that (\ref{SINR_Among_Group}) is guaranteed by (\ref{ValidSIC}). In addition, it is worth to point out that the minimum required SINRs $\{\Gamma_{g}\}_{g=1}^{G}$ is actually the target SINR levels of messages $\{s_g\}_{g=1}^{G}$ with the corresponding rates $\left\{\log_2\left(1+\Gamma_{g}\right)\right\}_{g=1}^{G}$ for the $G$ groups of users. In other words, under our proposed NOMA framework, the transmission rates of all the $G$ data streams are \textit{predefined} and \textit{fixed}. Accordingly, based on the predefined data rates, the minimum SINRs of being able to successfully decode the data streams can be figured out and are eventually interpreted as these thresholds $\{\Gamma_g\}_{g=1}^{G}$.

\section{Transmit Power Minimization Under Quality of Service Constraints}\label{SecQoS}
In this section, we focus on the problem of minimizing the total transmit power subject to an individual QoS constraint for each user. There are two categories of constraints in this optimization problem: one is the QoS constraints which guarantee the predefined target SINR levels for all users and the other is the SIC constraints which ensure the successful implementation of SIC given the predefined decoding order. Based on the signal model established in the previous section, this optimization problem can be formulated as
\begin{subequations}\label{Problem1}
\begin{align}
    &P_\textrm{Min}\triangleq\min_{\mathbf{w}_g, 1\leq g\leq G} \sum\nolimits_{g=1}^G\Big\|\mathbf{w}_g\Big\|^2_2\\
    \textrm{s.t.}~~
    &\textrm{SINR}_{u,g}\geq \Gamma_g,~1\leq g\leq G,~1\leq u\leq U_g,\label{QoS_Constraints}\\
    &\textrm{SINR}_{u,g}^i\geq \Gamma_{i},~2\leq g\leq G,~1\leq u\leq U_g,~1\leq i\leq g-1,\label{SIC_Constraints}
\end{align}
\end{subequations}
where $P_\textrm{Min}$ denotes the minimum required transmit power that satisfies the QoS requirements{\footnote{From the perspective of practical implementation, the QoS requirements should be bounded. Otherwise, the total transmit power may be too large to be feasible. Nevertheless, in this work, the optimization is supposed to be performed under a reasonable QoS requirements setting since its design is an interesting but challenging topic which is out of the scope of this work.}} for all users, (\ref{QoS_Constraints}) represents the $\sum_{g=1}^GU_g$ QoS constraints and (\ref{SIC_Constraints}) represents the $\sum_{g=2}^GU_g(g-1)$ SIC constraints. For example, consider that there are 3 regions ($G=3$) and there are 3 users in each region ($U_g=3,1\leq g\leq3$). Then there should be 9 QoS constraints and 9 SIC constraints. Particularly, there are several special cases of problem (\ref{Problem1}): when $G=1$, problem (\ref{Problem1}) degrades into a single-group multicast beamforming design problem, in which the application of NOMA is not needed; when $U_g=1$ for $1\leq g\leq G$, problem (\ref{Problem1}) degrades into a common broadcast beamforming design problem in a MISO NOMA system.

Unlike those transmit beamforming optimization problems considered in conventional multi-group multicast \cite{Luo_MultiGroupMulticast}, the extra constraints given in (\ref{SIC_Constraints}), i.e., the SIC constraints, are due to the application of NOMA. To further explore problem (\ref{Problem1}), we rewrite it by substituting $\textrm{SINR}_{u,g}$ and $\textrm{SINR}_{u,g}^i$ with the R.H.S. of (\ref{SINR_Each_Group}) and (\ref{SINR_Among_Group}), respectively. Thus, problem (\ref{Problem1}) is rewritten as
\begin{subequations}\label{Problem1.2}
	\begin{align}
	&P_\textrm{Min}\triangleq\min_{\mathbf{w}_g, 1\leq g\leq G} \sum\nolimits_{g=1}^G\Big\|\mathbf{w}_g\Big\|^2_2\\
	\textrm{s.t.}~~&
	\left|\mathbf{h}_{u,g}^H \mathbf{w}_g\right|^2\geq \Gamma_{g}\left(\sum\nolimits_{k=g+1}^G\left|\mathbf{h}_{u,g}^H \mathbf{w}_k\right|^2+\sigma^2\right),\nonumber\\
	&~~~~~~~~~~~~~~~~~~~~~~~~~~~~~1\leq g\leq G,~1\leq u\leq U_g,\label{ConsQoS}\\
	&\left|\mathbf{h}_{u,g}^H \mathbf{w}_i\right|^2\geq \Gamma_{i}\left(\sum\nolimits_{k=i+1}^G\left|\mathbf{h}_{u,g}^H \mathbf{w}_k\right|^2+\sigma^2\right),\nonumber\\
	&~~~~~~~~~~~~2\leq g\leq G,~1\leq u\leq U_g,~1\leq i\leq g-1.\label{ConsSIC}
	\end{align}
\end{subequations}
It can be easily verified that problem (\ref{Problem1.2}) is non-convex due to the quadratic terms at the left-hand sides (L.H.S.) of the inequality constraints in (\ref{ConsQoS}) and (\ref{ConsSIC}). In fact, problem (\ref{Problem1.2}) is NP-hard according to \cite{Luo_MultiGroupMulticast}. Therefore, it is difficult to solve this problem directly, which leads us to use convex approximation techniques to find an approximate solution for the original problem. In the following, an efficient algorithm is proposed to solve problem (\ref{Problem1.2}) based on the majorization and minimization (MM)\footnote{The MM is termed as minorization-maximization when the original problem is a maximization problem.}  algorithms\cite{MMA1,MMA2}.

\subsection{MM-based Method}\label{SPCA_Sec}
In this subsection, we develop a MM-based method to solve problem (\ref{Problem1.2}). The basic idea of the MM method is to substitute a simple optimization problem for a difficult one and iteration is the price for simplifying the original problem \cite{MMA1}. Specifically, in each iteration, every non-convex constraint is substituted with its inner convex approximation which serves as its tangent plane at a certain point. A local optimal solution is ensured by the MM method\cite{MMA2,SCA}. Thus, we are to analyze the non-convexity of problem (\ref{Problem1.2}) and find the approximation functions.

According to the formulation of problem (\ref{Problem1.2}), its main difficulty lies in the quadratic terms on the L.H.S. of the inequalities constraints in (\ref{ConsQoS}) and (\ref{ConsSIC}).
Noting that those quadratic terms are the squared norms of complex numbers, we decouple the complex numbers into their real parts and imaginary parts in order to unveil the hidden concavity or linearity. Consequently, an equivalent transformation of the quadratic terms on the L.H.S. of (\ref{ConsQoS}) is given as below:
\begin{subequations}
    \begin{align}
    &\left|\mathbf{h}_{u,g}^H \mathbf{w}_g\right|^2 = \left(\Phi_{u,g,g}^\textrm{R}\right)^2 + \left(\Phi_{u,g,g}^\textrm{I}\right)^2,\\
    &\Phi_{u,g,g}^\textrm{R} \triangleq \textrm{Re}\left\{\mathbf{h}_{u,g}^H \mathbf{w}_g\right\},\\
    &\Phi_{u,g,g}^\textrm{I} \triangleq \textrm{Im}\left\{\mathbf{h}_{u,g}^H \mathbf{w}_g\right\}.
    \end{align}
\end{subequations}
We further create a two-dimensional real-valued column vector given as
\begin{equation}
    \Phi_{u,g,g} \triangleq \left[\Phi_{u,g,g}^\textrm{R},\Phi_{u,g,g}^\textrm{I}\right]^T.
\end{equation}
Therefore, $\left|\mathbf{h}_{u,g}^H \mathbf{w}_g\right|^2$ is equivalently transformed to the squared norm of the created real-valued vector, which is shown as
\begin{equation}
    \left|\mathbf{h}_{u,g}^H \mathbf{w}_g\right|^2 = \left\|\Phi_{u,g,g}\right\|^2_2 \triangleq y\left(\Phi_{u,g,g}\right).\label{funcY}
\end{equation}
Since $y(\cdot)$ is the squared Euclidean norm of a real-valued vector, it follows that \cite{Convex_Optimization,Quadratic_Linea}
\begin{equation}\label{Taylor1}
\begin{split}
&y\left(\Phi_{u,g,g}\right)\geq y_T\left(\Phi_{u,g,g},\Phi_{u,g,g}^{\left(n\right)}\right)\triangleq \left\|\Phi_{u,g,g}^{\left(n\right)}\right\|_2^2+\\
&~~~~~~~~~~~~~~~~~~~~~~~~2\left(\Phi_{u,g,g}^{\left(n\right)}\right)^T\left(\Phi_{u,g,g}-\Phi_{u,g,g}^{\left(n\right)}\right),
\end{split}
\end{equation}
where $y_T\left(\Phi_{u,g,g},\Phi_{u,g,g}^{\left(n\right)}\right)$ is the first-order Taylor approximation of $y\left(\Phi_{u,g,g}\right)$ around $\Phi_{u,g,g}^{\left(n\right)}$, the setting of which will be given in the following. $y_T\left(\Phi_{u,g,g},\Phi_{u,g,g}^{\left(n\right)}\right)$ minorizes  $y\left(\Phi_{u,g,g}\right)$ at point $\Phi_{u,g,g}^{\left(n\right)}$, which means that $y_T\left(\Phi_{u,g,g},\Phi_{u,g,g}^{\left(n\right)}\right)$ lies under the surface of $y\left(\Phi_{u,g,g}\right)$ and is tangent to it at the point $\Phi_{u,g,g}=\Phi_{u,g,g}^{\left(n\right)}$. Thus, the linear expression $y_T\left(\Phi_{u,g,g},\Phi_{u,g,g}^{\left(n\right)}\right)$ can be used to approximate the original quadratic term $\left|\mathbf{h}_{u,g}^H \mathbf{w}_g\right|^2$. In a similar way, the quadratic terms in the L.H.S. of (\ref{ConsSIC}) can be also approximated as below:
\begin{subequations}\label{Taylor2}
	\begin{align}
	&\left|\mathbf{h}_{u,g}^H \mathbf{w}_i\right|^2=\left(\Phi_{u,g,i}^\textrm{R}\right)^2 + \left(\Phi_{u,g,i}^\textrm{I}\right)^2,\\
	&\Phi_{u,g,i}^\textrm{R} \triangleq \textrm{Re}\left\{\mathbf{h}_{u,g}^H \mathbf{w}_i\right\},\\
	&\Phi_{u,g,i}^\textrm{I} \triangleq \textrm{Im}\left\{\mathbf{h}_{u,g}^H \mathbf{w}_i\right\},\\
	&\Phi_{u,g,i} \triangleq \left[\Phi_{u,g,i}^\textrm{R},\Phi_{u,g,i}^\textrm{I}\right]^T,\\
	&\left|\mathbf{h}_{u,g}^H \mathbf{w}_i\right|^2 = \left\|\Phi_{u,g,i}\right\|_2^2 \triangleq g\left(\Phi_{u,g,i}\right),\\
	&g\left(\Phi_{u,g,i}\right) \geq g_T\left(\Phi_{u,g,i},\Phi_{u,g,i}^{\left(n\right)}\right)\triangleq\left\|\Phi_{u,g,i}^{\left(n\right)}\right\|_2^2+\nonumber\\
	&~~~~~~~~~~~~~~~~~~~~~~~~2\left(\Phi_{u,g,i}^{\left(n\right)}\right)^T\left(\Phi_{u,g,i}-\Phi_{u,g,i}^{\left(n\right)}\right).
	\end{align}
\end{subequations}
With the above approximations, the constraints in (\ref{ConsQoS}) and (\ref{ConsSIC}) can be approximated by more stringent but convex constraints, which makes the original problem (\ref{Problem1.2}) become an iterative convex program. Specifically, the $n$-th iteration is to solve the following convex optimization problem:
\begin{subequations}\label{Problem_Con_App_1n}
	\begin{align}
	&P_\textrm{Min}^\textrm{SCA}\triangleq\min_{\mathbf{w}_g, 1\leq g\leq G} \sum\nolimits_{g=1}^G\Big\|\mathbf{w}_g\Big\|^2_2\\
	\textrm{s.t.}~~&
	y_T\left(\Phi_{u,g,g},\Phi_{u,g,g}^{\left(n\right)}\right)\geq \Gamma_{g}\left(\sum\nolimits_{k=g+1}^G\left|\mathbf{h}_{u,g}^H \mathbf{w}_k\right|^2+\sigma^2\right),\nonumber\\
	&~~~~~~~~~~~~~~~~~~~~~~~~~~~1\leq g\leq G,~1\leq u\leq U_g,\label{ConsQoS2}\\
	&g_T\left(\Phi_{u,g,i},\Phi_{u,g,i}^{\left(n\right)}\right)\geq \Gamma_{i}\left(\sum\nolimits_{k=i+1}^G\left|\mathbf{h}_{u,g}^H \mathbf{w}_k\right|^2+\sigma^2\right),\nonumber\\
	&~~~~~~~~~2\leq g\leq G,~1\leq u\leq U_g,~1\leq i\leq g-1.\label{ConsSIC2}
	\end{align}
\end{subequations}
Here, the settings of $\Phi_{u,g,g}^{\left(n\right)}$ and $\Phi_{u,g,i}^{\left(n\right)}$ are given by
\begin{subequations}\label{IniPmin}
\begin{align}
&\Phi_{u,g,g}^{\left(n\right)}\triangleq\left[\textrm{Re}\left\{\mathbf{h}_{u,g}^H \mathbf{w}_g^{\left(n-1\right)}\right\},\textrm{Im}\left\{\mathbf{h}_{u,g}^H \mathbf{w}_g^{\left(n-1\right)}\right\}\right]^T,\\
&\Phi_{u,g,i}^{\left(n\right)}\triangleq\left[\textrm{Re}\left\{\mathbf{h}_{u,g}^H \mathbf{w}_i^{\left(n-1\right)}\right\},\textrm{Im}\left\{\mathbf{h}_{u,g}^H \mathbf{w}_i^{\left(n-1\right)}\right\}\right]^T,
\end{align}
\end{subequations}
where $\left\{\{\mathbf{w}_g^{\left(n\right)}\}_{g=1}^{G}\right\}$ denotes the optimal solution to problem (\ref{Problem_Con_App_1n}). Accordingly, the MM-based method for solving problem (\ref{Problem1.2}) is outlined in Algorithm 1. Moreover, the approach used to generate the initial feasible solution $\left\{\{\mathbf{w}_g^{\left(0\right)}\}_{g=1}^{G}\right\}$ and the convergence analysis of our proposed iterative algorithm will be discussed as follows, respectively.

\begin{algorithm}
\caption{Transmit Power Minimization Subject to QoS Constraints in NOMA Assisted Multi-Region Geocast}\label{alg:AlgPro1}
\begin{algorithmic}[1]
\REQUIRE Initial feasible solution $\left\{\{\mathbf{w}_g^{\left(0\right)}\}_{g=1}^{G}\right\}$, $\{\mathbf{h}_{u,g}\}_{u,g}$, $\sigma^2$, $\{\Gamma_{g}\}_{g=1}^G$;
%\ENSURE Suboptimal beamforming vectors to problem (\ref{Problem1.2});
\STATE $n=1$;
\STATE Initialize $\Phi_{u,g,g}^{\left(1\right)}$ and $\Phi_{u,g,i}^{\left(1\right)}$ with (\ref{IniPmin});
\REPEAT
\STATE Solve problem (\ref{Problem_Con_App_1n});
\STATE Update $n=n+1$;
\STATE Update $\Phi_{u,g,g}^{\left(n\right)}$ and $\Phi_{u,g,i}^{\left(n\right)}$ with (\ref{IniPmin});
\UNTIL{Convergence or limitation of the number of iterations;}
\end{algorithmic}
\end{algorithm}

\emph{Generation of Initial Feasible Solution}:
We now propose an efficient method to seek out an initial feasible solution for our proposed MM-based method. In fact, problem (\ref{Problem1.2}) can be approximated by a second-order cone program (SOCP), which is presented as follows\footnote{Another approach of generating an initial feasible solution for the general conic quadratic problems are also provided in \cite{Convergency}.}. We firstly rewrite the non-convex constraints in (\ref{ConsQoS}) as
\begin{equation}\label{ConsQoS_Rewrite}
\begin{split}
&\left|\mathbf{h}_{u,g}^H \mathbf{w}_g\right|\geq \sqrt{\Gamma_{g}\left(\sum\nolimits_{k=g+1}^G\left|\mathbf{h}_{u,g}^H \mathbf{w}_k\right|^2+\sigma^2\right)},\\
&~~~~~~~~~~~~~~~~~~~~~~1\leq g\leq G,~1\leq u\leq U_g.
\end{split}
\end{equation}
Further, the conservative approximation proposed in \cite{SOCPAPro} is applied as below:
\begin{equation}\label{socp_appromixation}
    \left|\mathbf{h}_{u,g}^H \mathbf{w}_g\right|\geq \textrm{Re}\left\{\mathbf{h}_{u,g}^H \mathbf{w}_g\right\}.
\end{equation}
By using (\ref{socp_appromixation}), the non-convex constraints in (\ref{ConsQoS_Rewrite}) are approximated by more stringent but convex constraints given by
\begin{equation}
\begin{split}
&\textrm{Re}\left\{\mathbf{h}_{u,g}^H \mathbf{w}_g\right\} \geq\sqrt{\Gamma_{g}\left(\sum\nolimits_{k=g+1}^G\left|\mathbf{h}_{u,g}^H \mathbf{w}_k\right|^2+\sigma^2\right)},\\
&~~~~~~~~~~~~~~~~~~~~~~~~~~~1\leq g\leq G,~1\leq u\leq U_g,
\end{split}
\end{equation}
which can be further rewritten under the second-order convex (SOC) representations given as
\begin{align}
&\textrm{Re}\left\{\mathbf{h}_{u,g}^H \mathbf{w}_g\right\} \geq\sqrt{\Gamma_{g}}
\begin{Vmatrix}
\left(\mathbf{h}_{u,g}^H \mathbf{w}_{g+1},
\ldots,
\mathbf{h}_{u,g}^H \mathbf{w}_{G},
\sqrt{\sigma^2}\right)
\end{Vmatrix}_2,\nonumber\\
&~~~~~~~~~~~~~~~~~~~~~~~~~~~~~~~~~~1\leq g\leq G,~1\leq u\leq U_g.\label{SOC_QoS}
\end{align}
In a similar way, the inner convex approximations of the constraints in (\ref{ConsSIC}) can be given by
\begin{align}
&\textrm{Re}\left\{\mathbf{h}_{u,g}^H \mathbf{w}_i\right\} \geq\sqrt{\Gamma_{i}}
\begin{Vmatrix}
\left(\mathbf{h}_{u,g}^H \mathbf{w}_{i+1},
\ldots,
\mathbf{h}_{u,g}^H \mathbf{w}_{G},
\sqrt{\sigma^2}\right)
\end{Vmatrix}_2,\nonumber\\
&~~~~~~~~~~~~2\leq g\leq G,~1\leq u\leq U_g,~1\leq i\leq g-1.\label{SOC_SIC}
\end{align}
As a result, problem (\ref{Problem1.2}) is approximated by the following standard SOCP:
\begin{subequations}\label{Problem1.SOC}
\begin{align}
    &\min_{\mathbf{w}_g, 1\leq g\leq G} \sum\nolimits_{g=1}^G\Big\|\mathbf{w}_g\Big\|^2_2\\
    \textrm{s.t.}~~&(\ref{SOC_QoS})~~\textrm{and}~~(\ref{SOC_SIC}).
\end{align}
\end{subequations}
Let us denote the optimal solution to problem (\ref{Problem1.SOC}) as $\left\{\{\mathbf{w}^{(0)}_g\}_{g=1}^G\right\}$ and set it as the initial feasible solution for Algorithm 1, then our proposed MM-based method can solve problem (\ref{Problem1.2}) in the vicinity of this initial solution $\left\{\{\mathbf{w}^{(0)}_g\}_{g=1}^G\right\}$ by finding a better solution. In particular, problem (\ref{Problem1.SOC}) can be easily solved with general interior-point methods\cite{Convex_Optimization} or the advanced SOCP solvers, e.g., MOSEK \cite{SOCPSolver1}, ECOS \cite{SOCPSolver2}.

\emph{Convergence Analysis}:
Following the general convergence principle of the MM method in \cite{MMA1,SCA}, we can prove the convergence of algorithm 1. According to \cite[Theorem 1]{SCA} and \cite{Convergency}, it can be verified that the first-order Taylor approximations in (\ref{Taylor1}) and (\ref{Taylor2}) satisfy the constraints for assuring the convergence of the MM method. As a result, it is assured that the solution obtained by Algorithm 1 converges to a Karush-Kuhn-Tucker solution of the original problem (\ref{Problem1.2}). Besides, our simulation results in Section \ref{SecSim} further show that Algorithm 1 converges fast in a finite number of iterations.

\emph{Complexity Analysis}:
In each iteration of Algorithm 1, the SOCP problem (\ref{Problem_Con_App_1n}) can be solved by interior-point methods, of which the complexity depends on the number of constraints \cite{ComplexityMM}. As mentioned before, there are $\sum_{g=1}^GU_g$ QoS constraints and $\sum_{g=2}^GU_g(g-1)$ SIC constraints in problem (\ref{Problem_Con_App_1n}). By setting $U_1=U_2=...=U_G=U$, the worst-case complexity for the SOCP problem (\ref{Problem_Con_App_1n}) is $\mathcal{O}(\sqrt{UG^2})$. This complexity scales linearly with the number of iterations, which is shown to be a small number by the simulation results provided in Section \ref{SecSim}.
%The overall complexity of the SDR approach is that of a single SDP problem along with Nrand times the complexity of a LP problem. Here, Nrand stands for the number of generated candidate vectors of which, according to [3], a few hundred are usually required. The mentioned SDP problem can be solved with a worst-case complexity of O(M2(GN +M)2.5) [8] whereas the LP requires a worst-case complexity of O(G3.5 + MG3.5) [3]. Thus, as increases, the computational complexity of our method decreases relative to that of the SDR-based method.

\subsection{Lower Bound of $P_\textrm{Min}$: Classical SDR Method}\label{SDR_Sec}
In this subsection, we use the classical SDR method to obtain the lower bound of the minimum required transmit power $P_\textrm{Min}$. Since problem (\ref{Problem1.2}) is a standard quadratically constrained quadratic programming (QCQP) problem, its suitable reformulation can realize the application of SDR. By introducing $\mathbf{X}_g=\mathbf{w}_g\mathbf{w}_g^H$ for $1\leq g\leq G$, problem (\ref{Problem1.2}) can be equivalently transformed as
\begin{subequations}\label{Problem_1SDR}
	\begin{align}
	&\min_{\mathbf{X}_g, 1\leq g\leq G} \sum\nolimits_{g=1}^G\textrm{Tr}\left(\mathbf{X}_g\right)\\
	\hspace{-0.07in}\textrm{s.t.}~
	&{\textrm{Tr}\left(\mathbf{h}_{u,g}^H\mathbf{X}_g\mathbf{h}_{u,g}\right)}\geq{\Gamma_{g}} \left[\sum\nolimits_{k=g+1}^G\textrm{Tr}\left(\mathbf{h}_{u,g}^H\mathbf{X}_k\mathbf{h}_{u,g}\right)+\sigma^2\right],\nonumber\\
	&~~~~~~~~~~~~~~~~~~~~~~~~~~~~1\leq g\leq G,~1\leq u\leq U_g,\\
	&{\textrm{Tr}\left(\mathbf{h}_{u,g}^H\mathbf{X}_i\mathbf{h}_{u,g}\right)}\geq{\Gamma_{i}} \left[\sum\nolimits_{k=i+1}^G\textrm{Tr}\left(\mathbf{h}_{u,g}^H\mathbf{X}_k\mathbf{h}_{u,g}\right)+\sigma^2\right],\nonumber\\
	&~~~~~~~~~~~2\leq g\leq G,~1\leq u\leq U_g,~1\leq i\leq g-1,\\
	&\mathbf{X}_g\succeq 0, ~1\leq g\leq G,\\
	&\textrm{rank}(\mathbf{X}_g) = 1,~1\leq g\leq G.\label{ConsRank}
	\end{align}
\end{subequations}
By dropping the rank-one constraints in (\ref{ConsRank}), which are the only existing non-convex constraints, problem (\ref{Problem_1SDR}) is thereby relaxed to a standard semi-definite programming (SDP) problem that can be solved by modern SDP solvers, such as SeDuMi\cite{Sedumi}. However, due to the relaxation, the obtained solution to the SDP problem may not be rank one. \footnote{The drawback of this relaxation method is that the solution of the SDP problem is not guaranteed to be feasible for its original problem. In some cases, the Gaussian randomization method (see, e.g., \cite{Luo_Multicast} and references therein) can generate a feasible approximate solution from the solution of the SDP problem.}
Thus, the optimal solution to the relaxed problem can serve as a lower bound of $P_\textrm{Min}$. In Section \ref{SecSim} where the numerical results are provided, we use this SDR method as a benchmark for our proposed MM-based method.

\section{System Sum Rate Maximization Under Fair Transmission Policy}\label{SecFair}
For many communication scenarios, the system sum rate is a more important criterion to evaluate the performance of spectrum sharing, compared with the transmit power. Therefore, in this section, based on the previous transmit power minimization problem, we further consider a related problem which is to maximize the achievable sum rate of the system subject to a total transmit power constraint. This investigation aims to explore the NOMA assisted multi-region geocast framework from the perspective of system throughput. In particular, a \emph{fair transmission policy} is proposed and adopted in this scenario, which is elaborated in the following:

We denote the target transmit data rate of the message for the $g$-th group as $R_g$ and meanwhile define $\{r_g\}_{g=1}^G$ as the prescribed grades of service for the $G$ groups (greater $r_g$ is, higher transmit data rate $R_g$ will be). Then, our proposed fair transmission policy requires that all ratios between the data rates of groups should be the ratios between their grades of service, which are mathematically characterized by the following equations:
\begin{equation}\label{FairPolicy}
    \frac{R_i}{R_j}=\frac{r_i}{r_j},~1\leq i\leq G,~1\leq j\leq G.
\end{equation}

Herein, ($r_1$, $r_2$, $...$, $r_G$) could be interpreted as a set of baseline data rates. Under the proposed fair transmission policy characterized by (\ref{FairPolicy}), we can further define $k$ as a scale coefficient of the grades of service, which means that $kr_g$ (bits/s/Hz) is set to be the new target transmit data rate $R_g$ of the message for the $g$-th group so that (\ref{FairPolicy}) is always satisfied. Accordingly, each target data rate will be multiplied by a factor $k$ and the newly desired set of rates becomes ($kr_1$, $kr_2$, $...$, $kr_G$) where $k$ could be either greater or less equal than 1. As a result, the minimum required SINR for successful decoding the message for $g$-th group, denoted by $T_g$, can be expressed as
\begin{equation}
     T_g = 2^{kr_{g}}-1,~1\leq g\leq G.
\end{equation}

Let $P_\textrm{Tot}$ represent the total transmit power available at the transmitter, then the problem of maximizing the achievable sum rate of the system, subject to the total transmit power constraint and under the predefined fair transmission policy, can be formulated as
\begin{subequations}\label{Problem2}
\begin{align}
    &R_\textrm{Sum}\triangleq\max_{k,\mathbf{w}_g, 1\leq g\leq G}~~k\sum\nolimits_{g=1}^{G}r_g\\
    \textrm{s.t.}~~&
    \textrm{SINR}_{u,g}\geq T_g,~1\leq g\leq G,~1\leq u\leq U_g, \label{P2QoS_Cons}\\
    &\textrm{SINR}_{u,g}^i\geq T_i,~2\leq g\leq G,~1\leq u\leq U_g,~1\leq i\leq g-1,\label{P2SIC_Cons}\\
    &\sum\nolimits_{g=1}^{G}\Big\|\mathbf{w}_g\Big\|^2_2\leq P_\textrm{Tot}.\label{Power_Cos}
\end{align}
\end{subequations}
The analysis and the solution of the above problem are discussed in the following.

\subsection{Bisection Method Combined with MM-based Method}
Comparing problem (\ref{Problem2}) with problem (\ref{Problem1}) studied in Section \ref{SecQoS}, we can see that the QoS and SIC constraints of both the problems preserve the same structure except that the R.H.S. of the constraints in (\ref{P2QoS_Cons}) and (\ref{P2SIC_Cons}) are not constants but the functions of argument $k$. Accordingly, a straightforward strategy to solve (\ref{Problem2}) is proposed as below:
by fixing $k$ to a constant $k^\textrm{Fix}$, we firstly solve problem (\ref{Problem1}) with $\Gamma_g=T_g^\textrm{Fix} = 2^{k^\textrm{Fix}r_{g}}-1$ by Algorithm 1. Then by comparing $P_\textrm{Min}^\textrm{MM}$ to $P_\textrm{Tot}$, we judge whether the fixed scale coefficient $k^\textrm{Fix}$ is feasible under the total transmit power constraint in (\ref{Power_Cos}). If it is, we further increase the value of $k^\textrm{Fix}$ to make the most use of the available power at the transmitter. If not, we should decrease the value of $k^\textrm{Fix}$ since $P_\textrm{Tot}$ cannot satisfy the current target transmit data rates of all user groups.

Based on the proposed strategy, the bisection method can be exploited to solve problem (\ref{Problem2}). To be more specific, the combination of the one-dimensional search on $k$ and the MM-based method can solve problem (\ref{Problem2}) approximately.

Due to the application of the bisection method, it is necessary to determine an appropriate searching range for $k$. Let $L^\textrm{k}$ and $U^\textrm{k}$ represent the lower bound and the upper bound of the searching range for $k$, respectively.
Apparently, $L^\textrm{k}$ should be initialized as 0.
As for $U^\textrm{k}$, assuming that the total transmit power is directed towards an arbitrary single user without regarding to the QoS guarantees of the other users, we can obviously find $\sum_{g=1}^GU_g$ inaccessible upper bounds for $k$, which are given by
\begin{equation}\label{Value_UpS}
\frac{\log_2\left(1+{\parallel\mathbf{h}_{u,g}\parallel_2^2P_\textrm{Tot}}/{\sigma^2}\right)}{r_g},~1\leq u\leq U_g,~1\leq g\leq G.
\end{equation}
In order to minimize the number of iterations of the bisection method, we choose the smallest one among (\ref{Value_UpS}) to be $U^\textrm{k}$, which is given by
\begin{equation}\label{Value_UpS_Tight}
U^\textrm{k} = \min\limits_{1\leq g\leq G}\frac{\log_2\left[1+{\min\limits_{1\leq u\leq U_g}\left({\parallel\mathbf{h}_{u,g}\parallel_2^2P_\textrm{Tot}}/{\sigma^2}\right)}\right]}{r_g}.
\end{equation}
As a result, the algorithm which combines the bisection method and the MM-based method proposed in Section \ref{SecQoS}, for solving problem (\ref{Problem2}) is outlined in Algorithm 2 where $\epsilon$ is the tolerance of the optimal scale coefficient $k^{\textrm{Opt}}$.

\begin{algorithm}
\caption{System Sum Rate Maximization Under Predefined Fair Transmission Policy in NOMA Assisted Multi-Region Geocast}\label{alg:AlgPro2}
\begin{algorithmic}[1]
\REQUIRE $\epsilon$, $P_\textrm{Tot}>0$, $\{\mathbf{h}_{u,g}\}_{u,g}$, $\{r_g\}_{g=1}^G$, $\sigma^2$;
%\ENSURE $k^{\textrm{Opt}}$, $\{\mathbf{w}^\textrm{Opt}_{g}\}_{g=1}^G$, $R_\textrm{Sum}$;
\STATE Set lower bound of the searching range for k: $L^\textrm{k} = 0$;
\STATE Set upper bound of the searching range for k: (\ref{Value_UpS_Tight});
\STATE $k^\textrm{Fix}=\frac{L^\textrm{k}+U^\textrm{k}}{2}$;
\STATE $dk=U^\textrm{k}-L^\textrm{k}$;
\WHILE{$dk>\epsilon$}
    \STATE Solve problem (\ref{Problem1}) by Algorithm 1 with $\Gamma_g=T_g^\textrm{Fix} = 2^{k^\textrm{Fix}r_{g}}-1$ for $1\leq g\leq G$;
    %The approximate minimum transmit power $P_\textrm{Min}^\textrm{MM}$ is obtained;
    \IF {$P_\textrm{Min}^\textrm{MM}<P_\textrm{Tot}$}
    \STATE ~~~~~~~~~~~~$L^\textrm{k} = k^\textrm{Fix}$;
    \ELSE
    \STATE ~~~~~~~~~~~~$U^\textrm{k} = k^\textrm{Fix}$;
    \ENDIF
    \STATE Update $k^\textrm{Fix}: k^\textrm{Fix}=\frac{L^\textrm{k}+U^\textrm{k}}{2}$;
    \STATE Update $dk: dk=U^\textrm{k}-L^\textrm{k}$;
\ENDWHILE
\STATE $k^{\textrm{Opt}}=k^\textrm{Fix}$;
\end{algorithmic}
\end{algorithm}

\section{Energy Efficiency Optimization of the Upgraded System With an Extra User Group}\label{SecEE}
As the 5G system must be more compatible and flexible, our proposed NOMA framework is supposed to provide multiple options for the wireless access under different criteria, allowing the users with better channel condition to access the channel dynamically is one of the most important features. Accordingly, in this subsection, we consider a scenario wherein an additional user group with delay-sensitive data locating in a region most closely to the transmitter intends to access the system, when there are already multiple user groups being served. 
This scenario may widely appear in 5G applications with delay sensitive requirements, such as vehicle-to-vehicle networking and Internet of Things applications. In this applications, there is no rigorous requirement on data rate but the energy efficiency of the system is a great concern due to the severe constraint on power supplement.

Based on the above discussions, in this section, an EE maximization problem is investigated in our proposed framework with an upgrade that an extra user group having better channel condition is simultaneously served. To be specific, the transmitter serves the original $G$ regions (groups) and the newly joined region simultaneously, subject to the constraint that the QoS is always guaranteed for the users in the original $G$ regions. The objective is to maximize the EE of the system. In particular, the EE is evaluated by the commonly adopted metric ``bits per Joule per Hertz (bits/Joule/Hz)'' \cite{EE1,EE2,EE3}, which is defined as the ratio of the system sum rate to the total power expenditure.

\subsection{Problem Formulation}
Based on the above discussion, the sum of the achievable data rate of the original $G$ user groups is a constant given by
\begin{equation}
R_\textrm{Ori} \triangleq \sum_{g=1}^{G}\log_2(1+\Gamma_\textrm{g}).
\end{equation}
With regard to the newly joined user group, its achievable data rate is restricted by the minimum SINR among its users, since a common message is transmitted to the users in the same group. We use $\textrm{G}_\textrm{New}$ as the index of the newly joined user group. Furthermore, since the $\textrm{G}_\textrm{New}$ region locates most closely to the transmitter, it is supposed to perform the SIC to eliminate all the inter-group interference, the achievable data rate of the $\textrm{G}_\textrm{New}$ can be expressed as
\begin{equation}
    R_\textrm{New} \triangleq \log_2\left[1+\min_{1\leq u\leq U_{\textrm{G}_\textrm{New}}}\left(\frac{\left|\mathbf{h}_{u,\textrm{G}_\textrm{New}}^H\mathbf{w}_{\textrm{G}_\textrm{New}}\right|^2}{\sigma^2}\right)\right].\\
\end{equation}
As a result, the EE of this upgraded multi-region geocast system can be given by
\begin{equation}\label{EEObj}
        \textrm{EE}\triangleq\frac{R_\textrm{Ori} + R_\textrm{new}}{P_t+P_c},
\end{equation}
where $P_t \triangleq \sum_{g=1}^G\parallel\mathbf{w}_g\parallel^2_2+\parallel\mathbf{w}_{\textrm{G}_\textrm{New}}\parallel^2_2$ is the total transmit power and $P_c$ represents the fixed power expenditure of the system. Consequently, the EE optimization problem is formulated as
\begin{subequations}\label{Problem3}
\begin{align}
    &\max_{\mathbf{w}_g, 1\leq g\leq G,\mathbf{w}_{\textrm{G}_\textrm{New}}}~~\textrm{EE}\label{EEObj2}\\
    \textrm{s.t.}~~
    &\textrm{\~{SINR}}_{u,g}\geq \Gamma_g,~1\leq g\leq G,~1\leq u\leq U_g,\label{QoS_Cons_Cog}\\
    &\textrm{\~{SINR}}_{u,g}^i\geq \Gamma_{i},~2\leq g\leq G,~1\leq u\leq U_g,~1\leq i\leq g-1,\label{SIC_Cons_Cog}\\
    &\textrm{SINR}_{u,{\textrm{G}_\textrm{New}}}^i\geq \Gamma_{i},~1\leq u\leq U_{\textrm{G}_\textrm{New}},~1\leq i\leq G,\label{SIC_Cons_Cog2}\\
    &\sum\nolimits_{g=1}^G\parallel\mathbf{w}_g\parallel^2_2+\parallel\mathbf{w}_{\textrm{G}_\textrm{New}}\parallel^2_2\leq P_\textrm{Tot},\label{Pow_Cons_Cog}
\end{align}
\end{subequations}
where
\begin{subequations}
\begin{align}
&\textrm{\~{SINR}}_{u,g}=\frac{\left|\mathbf{h}_{u,g}^H \mathbf{w}_g\right|^2}{\sum_{k=g+1}^G\left|\mathbf{h}_{u,g}^H \mathbf{w}_k\right|^2+\left|\mathbf{h}_{u,g}^H \mathbf{w}_{\textrm{G}_\textrm{New}}\right|^2+\sigma^2},\label{SINR_Each_Group_New}\\
&\textrm{\~{SINR}}_{u,g}^i=\frac{\left|\mathbf{h}_{u,g}^H \mathbf{w}_i\right|^2}{\sum_{k=i+1}^G\left|\mathbf{h}_{u,g}^H \mathbf{w}_k\right|^2+\left|\mathbf{h}_{u,g}^H \mathbf{w}_{\textrm{G}_\textrm{New}}\right|^2+\sigma^2},\label{SINR_Among_Group_New}\\
&\textrm{SINR}_{u,{\textrm{G}_\textrm{New}}}^i=\frac{\left|\mathbf{h}_{u,{\textrm{G}_\textrm{New}}}^H \mathbf{w}_i\right|^2}{\sum_{k=i+1}^G\left|\mathbf{h}_{u,{\textrm{G}_\textrm{New}}}^H \mathbf{w}_k\right|^2+\sigma^2}\label{SINR_New3},
\end{align}
\end{subequations}
where (\ref{SINR_Each_Group_New}) and (\ref{SINR_Among_Group_New}) are obtained by taking the newly joined region $\textrm{G}_\textrm{New}$ into consideration and revising (\ref{SINR_Each_Group}) and (\ref{SINR_Among_Group}), respectively.

The constraints in (\ref{QoS_Cons_Cog}) and (\ref{SIC_Cons_Cog}) guarantee the minimum required QoS of the users in the original $G$ user groups and its implementation of SIC, respectively, the constraints in (\ref{SIC_Cons_Cog}) ensure that the users in the newly joined group can successful perform SIC to eliminate all the inter-group interference, and the constraint in (\ref{Pow_Cons_Cog}) is the total transmit power constraint. Different from the previous optimization problems (\ref{Problem1}) and (\ref{Problem2}), the difficulty of problem (\ref{Problem3}) lies in its complicated objective function that is fractional. Most of the current solutions to maximize a fractional expression are turning to the Dinkelbach's method\cite{Dinkelbach}, the computational complexity of which could be very high. Hereinafter, an efficient beamforming design algorithm is proposed by using the MM method.

\subsection{MM-based Method for EE Maximization}
Note that the approximation methods of the constraints in (\ref{QoS_Cons_Cog}), (\ref{SIC_Cons_Cog}), and (\ref{SIC_Cons_Cog2}) have already been studied in Subsection \ref{SPCA_Sec} by linearizing quadratic terms, thus the difficulty of resolving the problem only rests with its fractional objective in (\ref{EEObj2}). We decompose (\ref{EEObj2}) by introducing some auxiliary variables to further expose its hidden concavity. Then, problem (\ref{Problem3}) is rewritten as
\begin{subequations}\label{Problem3.1}
\begin{align}
    &\max_{\mathbf{w}_g, 1\leq g\leq G,\mathbf{w}_{\textrm{G}_\textrm{New}}}~~t\\
    \textrm{s.t.}~~& R_\textrm{Ori}+R_\textrm{New}\geq \sqrt{tz},\label{fenzi}\\
    &P_t+P_c \leq \sqrt{z}\label{fenmu},\\
    &(\textrm{\ref{QoS_Cons_Cog}})~,(\textrm{\ref{SIC_Cons_Cog}})~\textrm{and}~(\textrm{\ref{SIC_Cons_Cog2}}),\\
    &(\textrm{\ref{Pow_Cons_Cog}})\label{Pow_Cons_Cog2},
\end{align}
\end{subequations}
where $z$ and $t$ are the auxiliary variables which could be explained as the squared power consumption and the squared EE, respectively. The equivalence between (\ref{Problem3}) and (\ref{Problem3.1}) is ensured by that the constraints in (\ref{fenzi}) and (\ref{fenmu}) should hold with equality at optimum, and that maximizing $\sqrt{t}$ is equivalent to maximizing $t$.
Note that (\ref{fenmu}) is convex while (\ref{fenzi}) is non-convex due to the bilinear product term $\sqrt{tz}$. We equivalently transform (\ref{fenzi}) as below:
\begin{subequations}
    \begin{align}
        &\frac{\left|\mathbf{h}_{u,{\textrm{G}_\textrm{New}}}^H\mathbf{w}_{\textrm{G}_\textrm{New}}\right|^2}{\sigma^2}\geq v,~1\leq u\leq U_{\textrm{G}_\textrm{New}},\\
        &1+v\geq2^r, \label{vr}\\
        &R_\textrm{Ori}+r \geq \sqrt{tz},
    \end{align}
\end{subequations}
where $v$ is the auxiliary variable which can be interpreted as an attainable SINR level for the users in the group ${\textrm{G}_\textrm{New}}$ to decode their target message, and $r$ is the auxiliary variable which is interpreted as the data rate of $v$, namely the achievable rate $R_\textrm{New}$. Then, problem (\ref{Problem3.1}) has been equivalently transformed into
\begin{subequations}\label{Problem3.2}
\begin{align}
    &\max_{\mathbf{w}_g, 1\leq g\leq G,\mathbf{w}_{\textrm{G}_\textrm{New}}}~~t\\
    \textrm{s.t.}~~
    &\frac{\left|\mathbf{h}_{u,{\textrm{G}_\textrm{New}}}^H\mathbf{w}_{\textrm{G}_\textrm{New}}\right|^2}{\sigma^2}\geq v,~1\leq u\leq U_{\textrm{G}_\textrm{New}},\label{C1}\\
    &R_\textrm{Ori}+r \geq \sqrt{tz},\label{Ctz}\\
    &(\textrm{\ref{QoS_Cons_Cog}})~,(\textrm{\ref{SIC_Cons_Cog}})~\textrm{and}~(\textrm{\ref{SIC_Cons_Cog2}}),\label{CSICQoS}\\
    &(\textrm{\ref{fenmu}}),~(\textrm{\ref{Pow_Cons_Cog2}})~\textrm{and}~(\textrm{\ref{vr}}) \label{Convex},
\end{align}
\end{subequations}
where the constraints in (\ref{Convex}) are all the convex ones while those in (\ref{C1}), (\ref{Ctz}) and (\ref{CSICQoS}) are all non-convex. However, the constraints in (\ref{C1}) and (\ref{CSICQoS}) can be approximated by certain convex ones with the approximation methods proposed in Subsection \ref{SPCA_Sec}, e.g., the approximations given in (\ref{Taylor1}) and (\ref{Taylor2}). Thereby, (\ref{Ctz}) is the only remaining non-convex constraint that needs to be addressed.
We note that $\sqrt{tz}$ is jointly concave in relation to $t$ and $z$ when $t>0$ and $z>0$. Consequently, an upper convex surface plane for $\sqrt{tz}$ could be given as below\cite{FractionMM}:
\begin{subequations}\label{Sqrttz}
	\begin{align}
	&\sqrt{tz} \triangleq f(t,z) \leq f_T\left(t,z,t^{\left(n\right)},z^{\left(n\right)}\right),\\
	&f_T\left(t,z,t^{\left(n\right)},z^{\left(n\right)}\right) \triangleq \sqrt{t^{\left(n\right)}z^{\left(n\right)}}+\frac{1}{2}\sqrt{\frac{z^{\left(n\right)}}{t^{\left(n\right)}}}\left(t-t^{\left(n\right)}\right)\nonumber\\
	&~~~~~~~~~~~~~~~~~~~~~~~~~~~+\frac{1}{2}\sqrt{\frac{t^{\left(n\right)}}{z^{\left(n\right)}}}\left(z-z^{\left(n\right)}\right),
	\end{align}
\end{subequations}
%,\\
%    &f_T\left(t,z,t^{\left(n\right)},z^{\left(n\right)}\right)
where $f_T\left(t,z,t^{\left(n\right)},z^{\left(n\right)}\right)$ is the first-order Taylor approximation of $f(t,z)$ around $\left(t^{\left(n\right)},z^{\left(n\right)}\right)$ whose values are to be set in the following. With (\ref{Sqrttz}), the non-convex constraint in (\ref{Ctz}) can be approximated by a more stringent but convex constraint shown as
\begin{equation}\label{CtzMM}
    R_\textrm{Ori}+r \geq f_T\left(t,z,t^{\left(n\right)},z^{\left(n\right)}\right).
\end{equation}
As a result, by using the convex approximation methods proposed in (\ref{Taylor1}), (\ref{Taylor2}) and (\ref{CtzMM}), the original problem (\ref{Problem3}) can be approximated as a convex problem in iterations. To be specific, the $n$-th iteration is to solve the following convex problem:
\begin{subequations}\label{Problem_Con_App_3n}
\begin{align}
    &\max_{\mathbf{w}_g, 1\leq g\leq G,\mathbf{w}_{\textrm{G}_\textrm{New}}}~~t\\
    \textrm{s.t.}~~
    &\frac{y_T\left(\Phi_{u,{\textrm{G}_\textrm{New}},{\textrm{G}_\textrm{New}}},\Phi_{u,{\textrm{G}_\textrm{New}},{\textrm{G}_\textrm{New}}}^{\left(n\right)}\right)}{\sigma^2}\geq v,1\leq u\leq U_{\textrm{G}_\textrm{New}},\label{C1P3}\\
    &y_T\left(\Phi_{u,g,g},\Phi_{u,g,g}^{\left(n\right)}\right)\nonumber\\
    &~~~\geq\Gamma_{g}\left(\sum\nolimits_{k=g+1}^G\left|\mathbf{h}_{u,g}^H \mathbf{w}_k\right|^2+\left|\mathbf{h}_{u,g}^H \mathbf{w}_{\textrm{G}_\textrm{New}}\right|^2+\sigma^2\right),\nonumber\\
    &~~~~~~~1\leq g\leq G,~1\leq u\leq U_g,\label{ConsQoSP3}\\
    &g_T\left(\Phi_{u,g,i},\Phi_{u,g,i}^{\left(n\right)}\right)\nonumber\\
    &~~~\geq\Gamma_{i}\left(\sum\nolimits_{k=i+1}^G\left|\mathbf{h}_{u,g}^H \mathbf{w}_k\right|^2+\left|\mathbf{h}_{u,g}^H \mathbf{w}_{\textrm{G}_\textrm{New}}\right|^2+\sigma^2\right),\nonumber\\
    &~~~~~~~2\leq g\leq G,~1\leq u\leq U_g,~1\leq i\leq g-1.\label{ConsSICP3}\\
    &(\textrm{\ref{fenmu}}),~(\textrm{\ref{Pow_Cons_Cog2}}),~(\textrm{\ref{vr}})~\textrm{and}~(\textrm{\ref{CtzMM}}).
\end{align}
\end{subequations}
Herein, the settings of $\Phi_{u,{\textrm{G}_\textrm{New}},{\textrm{G}_\textrm{New}}}^{\left(n\right)}$, $\Phi_{u,g,g}^{\left(n\right)}$ and $\Phi_{u,g,i}^{\left(n\right)}$ are similar to those given in (\ref{IniPmin}) while $z^{\left(n\right)}$ and $t^{\left(n\right)}$ are given by
\begin{subequations}\label{IniCog}
\begin{align}
&z^{\left(n\right)} = \left(\parallel\mathbf{w}_1^{\left(n-1\right)}\parallel^2_2+\parallel\mathbf{w}_2^{\left(n-1\right)}\parallel^2_2+P_C\right)^{2},\\
&t^{\left(n\right)} = \frac{\left\{R_\textrm{Ori}+\log_2\left[1+\min\limits_{1\leq u\leq U_{\textrm{G}_\textrm{New}}}\left(\frac{\left|\mathbf{h}_{u,{\textrm{G}_\textrm{New}}}^H\mathbf{w}_{\textrm{G}_\textrm{New}}^{\left(n-1\right)}\right|^2}{\sigma^2}\right)\right]\right\}^2}{z^{\left(n\right)}},
\end{align}
\end{subequations}
where $\left\{\{\mathbf{w}_g^{\left(n\right)}\}_{g=1}^{G},\mathbf{w}^{\left(n\right)}_{\textrm{G}_\textrm{New}}\right\}$ represents the optimal solution to problem (\ref{Problem_Con_App_3n}). Accordingly, the MM-based method for solving problem (\ref{Problem3}) is outlined in Algorithm 3. Besides, the initial feasible solution for the proposed iterative algorithm is $\left\{\{\mathbf{w}_g^{\left(0\right)}\}_{g=1}^{G},\mathbf{w}^{\left(0\right)}_{\textrm{G}_\textrm{New}}\right\}$, of which the generation is similar to the SOCP approximation method used in Section \ref{SecQoS}. In addition, the convergence analysis and the complexity analysis of Algorithm 3 are similar to those of Algorithm 1.
\begin{algorithm}
\caption{Energy Efficiency Maximization in the Updated NOMA Assisted Multi-Region Geocast System}\label{alg:AlgPro3}
\begin{algorithmic}[1]
\REQUIRE Initial feasible solution $\left\{\{\mathbf{w}_g^{\left(0\right)}\}_{g=1}^{G},\mathbf{w}^{\left(0\right)}_{\textrm{G}_\textrm{New}}\right\}$, $\{\mathbf{h}_{u,g}\}_{u,g}$, $\{\mathbf{h}_{u,{\textrm{G}_\textrm{New}}}\}_{u=1}^{U_{\textrm{G}_\textrm{New}}}$, $\Gamma_\textrm{Pri}$, $P_c$, $P_\textrm{Tot}$, $\sigma^2$;
%\ENSURE Suboptimal beamforming vectors to problem (\ref{Problem3});
\STATE $n=1$;
\STATE Initialize $\Phi_{u,{\textrm{G}_\textrm{New}},{\textrm{G}_\textrm{New}}}^{\left(1\right)}$, $\Phi_{u,g,g}^{\left(1\right)}$, $\Phi_{u,g,i}^{\left(1\right)}$, $z^{\left(1\right)}$ and $t^{\left(1\right)}$ with (\ref{IniPmin}) and (\ref{IniCog});
\REPEAT
\STATE Solve problem (\ref{Problem_Con_App_3n});
\STATE Update $n=n+1$;
\STATE Update  $\Phi_{u,{\textrm{G}_\textrm{New}},{\textrm{G}_\textrm{New}}}^{\left(n\right)}$, $\Phi_{u,g,g}^{\left(n\right)}$, $\Phi_{u,g,i}^{\left(n\right)}$, $z^{\left(n\right)}$ and $t^{\left(n\right)}$ with (\ref{IniPmin}) and (\ref{IniCog});
\UNTIL{Convergence or limitation of the number of iterations;}
\end{algorithmic}
\end{algorithm}

\section{Numerical Results}\label{SecSim}
In this section, we numerically evaluate the average performance of our proposed NOMA assisted multi-region geocast framework. For comparison, simulation results for two benchmarking schemes are also provided, which are described in the following. One is conventional multi-group multicast using SDMA\cite{Luo_MultiGroupMulticast}, in which the users in any group directly decode their own desired messages by treating the messages for the other groups as noise. The other one is single-group multicast using OMA, in which each group's desired message is transmitted in different orthogonal channel uses, such as time slots (in this work), to avoid the inter-group interference.
For notational simplicity, we use \emph{NOMA Scheme}, \emph{SDMA Scheme} and \emph{OMA Scheme} to refer to, respectively, NOMA assisted multi-region geocast, conventional multi-group multicast using SDMA and single-group multicast using OMA.
In particular, beamforming designs for both SDMA Scheme and OMA Scheme can be addressed by similar algorithms to the ones proposed in this work with NOMA. Besides, the lower bound of $P_\textrm{Min}$ for the first scenario and the upper bound of $R_\textrm{Sum}$ for the second scenario are provided by using the classical SDR method discussed in Section \ref{SDR_Sec}. Regarding to the upper bound of EE for the third scenario, the combination of the SDR method and the one-dimensional search on the total transmit power $P_\textrm{Tot}$ can help to obtain the upper bound of EE for the three aforementioned transmission schemes.

In each simulation, the problem is solved for 150 times using randomly generated CSI. We set the number of users in each group to the same value, denoted by $U$, namely $U_1=U_2=...U_G=U_{\textrm{G}_{\textrm{New}}}=U$. The specific values of parameters will be given under each figure of simulation results.
For the first scenario, we set the minimum required SINR for successfully decoding any message to the same value, denoted by $\Gamma$, thus the prescribed data rate of any group is $R_\Gamma = \log_2(1+\Gamma)$.
For the second scenario, we set each group's grade of service to the same level, namely $r_1=r_2=...r_G$, which means that all group's data rates are the same.
For the third scenario, our algorithm that maximizes the EE of the system is labeled as ``EEmax''. Another strategy that using full power $P_\textrm{Tot}$ to maximize the sum rate of the system, i.e., $R_\textrm{Ori} + R_\textrm{Sec}$, is presented for benchmarking, which is labeled as ``SEmax''.

\begin{figure}[!t]
    \centering
    \includegraphics[scale=0.30]{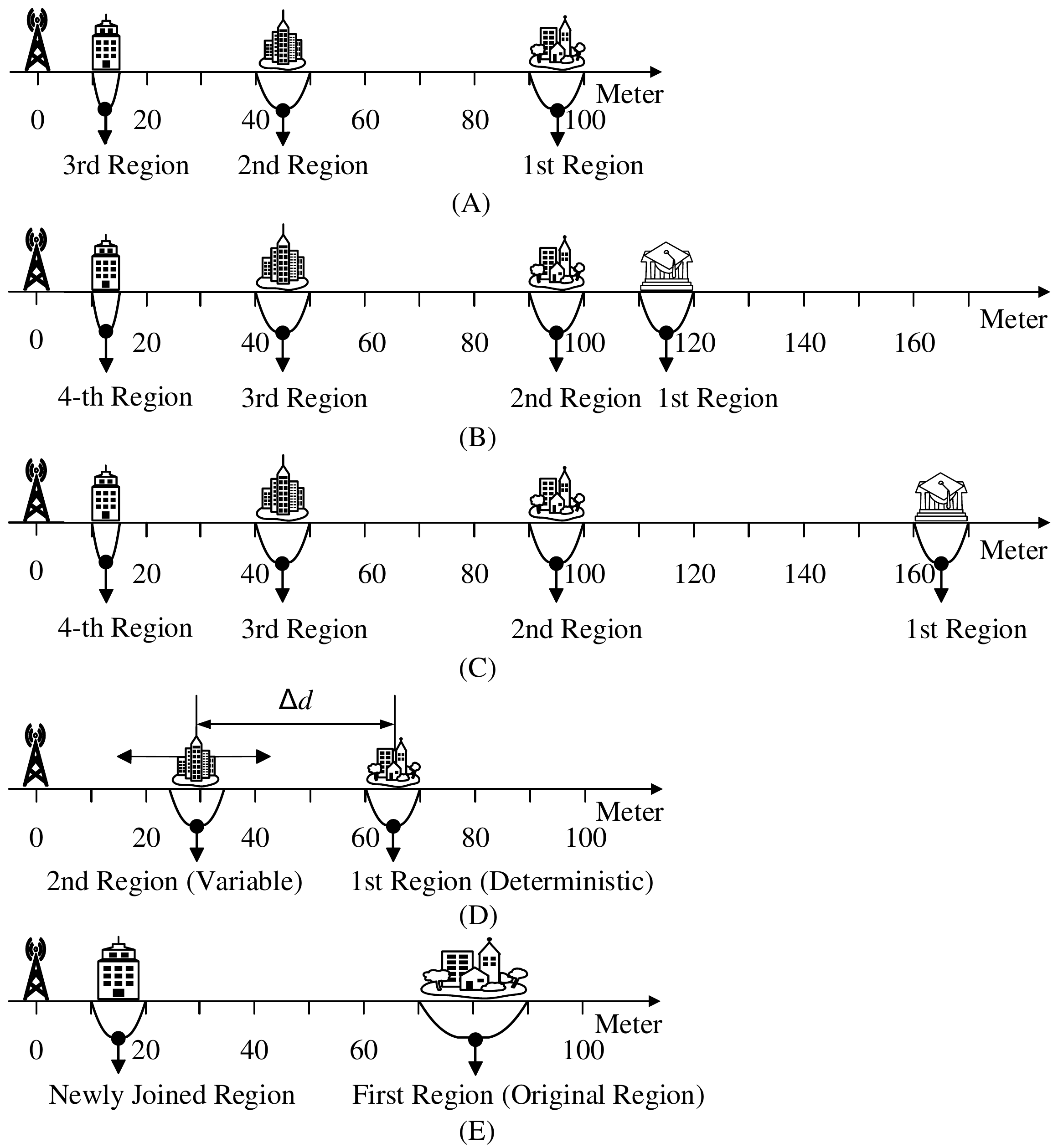}
    \caption{Four region location models are used in our numerical experiments:
    \protect\\(A) depicts a deterministic region location model, in which there are three regions with the parameters $d_1^\textrm{min}=90$m, $d_1^\textrm{max}=100$m, $d_2^\textrm{min}=40$m, $d_2^\textrm{max}=50$m, $d_3^\textrm{min}=10$m and $d_3^\textrm{max}=15$m.
    \protect\\(B) depicts a deterministic region location model, in which there are four regions with the parameters $d_1^\textrm{min}=110$, $d_1^\textrm{max}=120$m, $d_2^\textrm{min}=90$m, $d_2^\textrm{max}=100$m, $d_3^\textrm{min}=40$m, $d_3^\textrm{max}=50$m, $d_4^\textrm{min}=10$m and $d_4^\textrm{max}=15$m.
    \protect\\(C) depicts a deterministic region location model, in which there are four regions with the parameters $d_1^\textrm{min}=160$, $d_1^\textrm{max}=170$m, $d_2^\textrm{min}=90$m, $d_2^\textrm{max}=100$m, $d_3^\textrm{min}=40$m, $d_3^\textrm{max}=50$m, $d_4^\textrm{min}=10$m and $d_4^\textrm{max}=15$m.
    \protect\\(D) depicts a variable region location model consisting of two regions. The farther region (1st region) is deterministic with the parameters $d_1^\textrm{min}=60$m and $d_1^\textrm{max}=70$m while the nearer one (2nd region) is variable with the parameters $d_2^\textrm{min}=60-\Delta d$ m and $d_2^\textrm{max}=70-\Delta d$ m, where $\Delta d$ is the difference of their average distances to the transmitter. This variable region location model aims to explore the impact of the disparity level between the regions on the application of NOMA in multi-region geocast.
    \protect\\(E) depicts a deterministic region location model, in which there are two regions with the parameters $d_1^\textrm{min}=70$m, $d_1^\textrm{max}=90$m, $d_2^\textrm{min}=10$m, $d_2^\textrm{max}=20$m.}
    \label{fig:LocationModel}
\end{figure}

Regarding to region location models, three deterministic region location models and one variable region location model are used in our simulations, which are detailed in Fig. \ref{fig:LocationModel}.

\begin{figure}[!t]
    \centering
    \includegraphics[height = 6.9cm, width = 8.4cm]{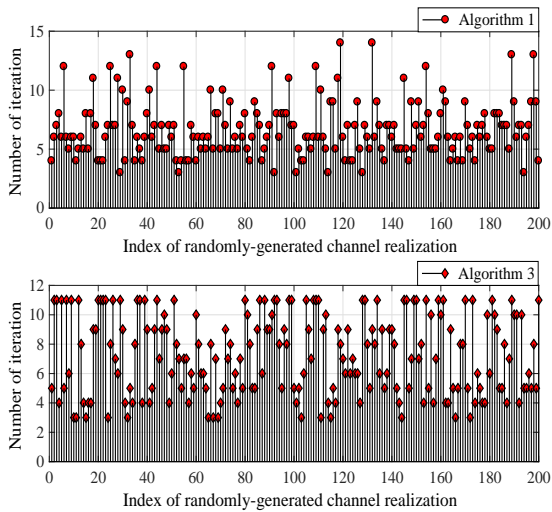}
    %\vspace{-12pt}
    \caption{Iterations required by Algorithm 1 and 3 for 200 randomly-generated channel realizations. Parameters for Algorithm 1: $M=8$, $G=3$, $U=3$, $R_\Gamma=3$ bits/s/Hz, $\alpha=2.5$ and $\sigma^2=-80$ dBm. Parameters for Algorithm 3: $R_\textrm{Pri}=1$ bits/s/Hz, $M=6$, $G=2$, $U=5$, $P_\textrm{Tot}=30$ dBm, $\alpha=2.5$, $\sigma^2=-80$ dBm and $P_c=44.77$ dBm (30 Watt).}
    \label{fig:NoI}
    %\vspace{-12pt}
\end{figure}
For the sake of illustrating the convergence rate of the proposed algorithms, we plot the convergence rates of Algorithm 1 and 3 for 200 randomly-generated channel realizations in Fig. \ref{fig:NoI}. It is worth pointing out that Algorithm 2 is based on Algorithm 1 in each step of the bisection method. Accordingly, the convergence rate of Algorithm 2 is omitted here. In these 200 channel realizations, we find that the maximum number of iterations for Algorithm 1 and 3 are 14 and 11, respectively, and for most cases, Algorithm 1 and 3 both converge within 10 iterations. As a result, we conclude that our proposed algorithms converge fast, which validates the practicality of the proposed schemes for enhancing the spectrum sharing.

\subsection{Simulation Results for the First Scenario}
\begin{figure}[!t]
    \centering
    \includegraphics[height = 6.9cm, width = 8.4cm]{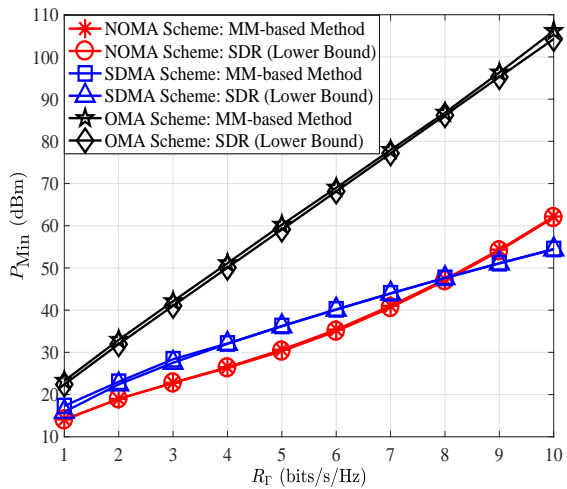}
    %\vspace{-12pt}
    \caption{Average minimum transmit power $P_\textrm{Min}$ (dBm) versus prescribed data rate $R_\Gamma$ (bits/s/Hz) for different transmission schemes. Parameters: $M=8$, $G=3$, $U=3$, $\alpha=2.5$ and $\sigma^2=-80$ dBm.}
    \label{fig:PMIN_R}
    %\vspace{-12pt}
\end{figure}
Fig. \ref{fig:PMIN_R} depicts average minimum transmit power $P_\textrm{Min}$ versus prescribed data rate $R_\Gamma$, under the region location model shown in Fig. \ref{fig:LocationModel}(A). Three interesting phenomena are summarized as below:
1) firstly, we can see that the curve generated by our proposed MM-based method is very close to the lower bound of $P_\textrm{Min}$ obtained by the SDR method, which indicates that our MM-based method has a high level of accuracy with acceptable approximate errors;
2) secondly, NOMA Scheme has a significant performance improvement compared with OMA Scheme. Even though no inter-group interference exists in OMA Scheme, it still consumes much more power than the other two schemes. This is because, in OMA Scheme, each group only occupies a part of time resources, which consequently results in much more power consumption for attaining the prescribed data rate in a limited time slot;
3) last but more importantly, NOMA Scheme does not always outperform SDMA Scheme. We can see that NOMA Scheme has a superior performance when $R_\Gamma$ is not too large. This is because the cost of SIC is relatively small when $R_\Gamma$ is small and the near-far effect, caused by different distances of the regions to the transmitter, can significantly improve system performance thanks to the elimination of partial co-channel interference. When $R_\Gamma$ becomes larger, the cost of SIC will increase since the nearer regions need more power to perform SIC, which degrades the performance of NOMA accordingly.

\begin{figure}[!t]
    \centering
    \includegraphics[height = 6.9cm, width = 8.4cm]{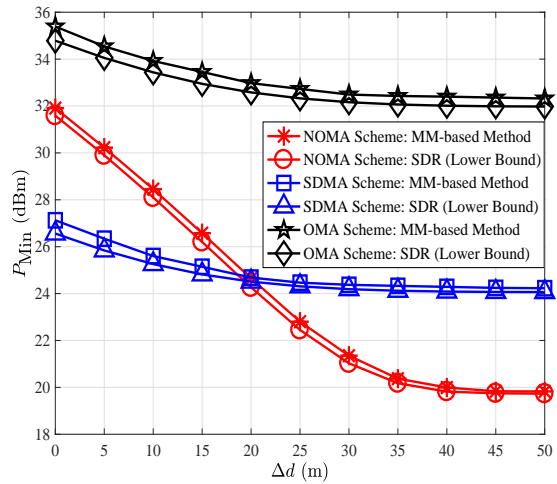}
    %\vspace{-12pt}
    \caption{Average minimum transmit power $P_\textrm{Min}$ (dBm) versus $\Delta d$ (m), where $\Delta d$ is defined as the difference of the two regions' average distances to the transmitter. Parameters: $R_\Gamma=4$ bits/s/Hz, $M=6$, $G=2$, $U=4$, $\alpha=2.5$ and $\sigma^2=-80$ dBm.}
    \label{fig:PMIN_DeltaDistance}
    %\vspace{-12pt}
\end{figure}
Fig. \ref{fig:PMIN_DeltaDistance} investigates the impact of region locations on the performance of NOMA by using the variable region location model shown in Fig. \ref{fig:LocationModel}(D). Firstly, we can see that when $\Delta d$ is not large enough, which implies that there is no remarkable channel gain difference between the users in two different regions, the performance gain of NOMA Scheme over SDMA Scheme is insignificant. Secondly, as expected, NOMA Scheme outperforms both SDMA Scheme and OMA Scheme when $\Delta d$ is sufficiently large. This is because as $\Delta d$ becomes larger, the near-far effect becomes more significant and dominant than the cost of SIC, which helps NOMA Scheme to achieve superior performance.

In the following three numerical experiments, we further investigate the pros and cons of NOMA Scheme and SDMA Scheme by applying different system parameters. For clearness and simplicity, the curves of the lower bound of $P_\textrm{Min}$ are omitted.

\begin{figure}[!t]
    \centering
    \includegraphics[height = 6.9cm, width = 8.4cm]{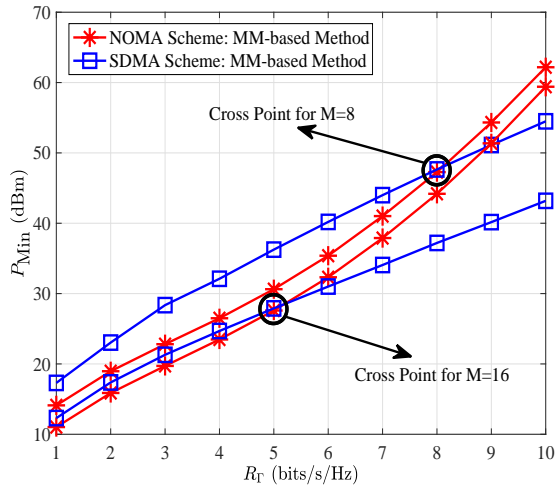}
    %\vspace{-12pt}
    \caption{Average minimum transmit power $P_\textrm{Min}$ (dBm) versus prescribed data rate $R_\Gamma$ (bits/s/Hz) for different numbers of transmit antennas $M$. Parameters: $G=3$, $U=3$, $\alpha=2.5$ and $\sigma^2=-80$ dBm.}
    \label{fig:PMIN_R_M}
    %\vspace{-12pt}
\end{figure}
Fig. \ref{fig:PMIN_R_M} compares the performance of NOMA Scheme with that of SDMA Scheme for different numbers of transmit antennas $M$, under the region location model shown in Fig. \ref{fig:LocationModel}(A). First of all, we can see that as $M$ increases, the power consumption of both schemes decreases due to the array gains provided by adding more transmit antennas.
However, the performance improvement of SDMA Scheme is more significant than that of NOMA Scheme.
This is because the increase of $M$ not only provides more degrees of freedom for naturally reducing the inter-group interferences, but also makes the channel vectors of the users tend to be orthogonal\footnote{Under the channel model given in (\ref{channel_gain}), the expectation of the inner product of two arbitrary users' channel vectors approaches zero as $M$ tends to infinity.}, which makes SIC become harder to implement and higher power-consuming, consequently restricting the performance improvement of NOMA Scheme.
Furthermore, it can be seen that there is a cross point for the curves generated by the two schemes, which indicates that the two transmission schemes are complementary and should be used as a combination in practice. Particularly, the cross point is left shifted as $M$ increases, which indicates that the range ensuring that NOMA Scheme is superior to SDMA Scheme is reduced. These observations imply that NOMA Scheme is more beneficial when the number of transmit antennas is limited.

\begin{figure}[!t]
    \centering
    \includegraphics[height = 6.9cm, width = 8.4cm]{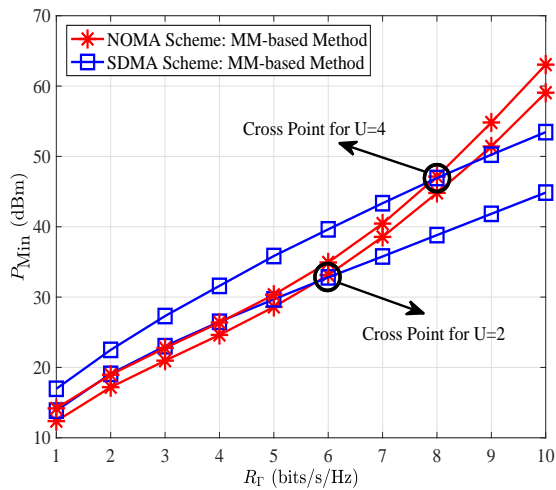}
    %\vspace{-12pt}
    \caption{Average minimum transmit power $P_\textrm{Min}$ (dBm) versus prescribed data rate $R_\Gamma$ (bits/s/Hz) for different numbers of users in a group $U$. Parameters: $M=10$, $G=3$, $\alpha=2.5$ and $\sigma^2=-80$ dBm.}
    \label{fig:PMIN_R_U}
    %\vspace{-12pt}
\end{figure}
Fig. \ref{fig:PMIN_R_U} further provides the performance comparison between NOMA Scheme and SDMA Scheme from the perspective of user numbers in each group, under the region location model shown in Fig. \ref{fig:LocationModel}(A). As $U$ increases, it is obvious that the system needs more power to guarantee the prescribed data rate for each user by using either transmission scheme. More importantly, we can see that the increase of $U$ has a smaller impact on NOMA Scheme than on SDMA Scheme. This is because NOMA Scheme exploits controllable inter-group interference to support massive users at the cost of a tolerable increase in the complexity of SIC, which makes it less sensitive to the number of users. On the contrary, SDMA Scheme simply treats the co-channel interference as noise, which consequently makes interference-resistance more difficult as $U$ increases. Accordingly, we conclude that NOMA Scheme is more favorable and robust for the massive connectivity scenario.

\begin{figure}[!t]
    \centering
    \includegraphics[height = 6.9cm, width = 8.4cm]{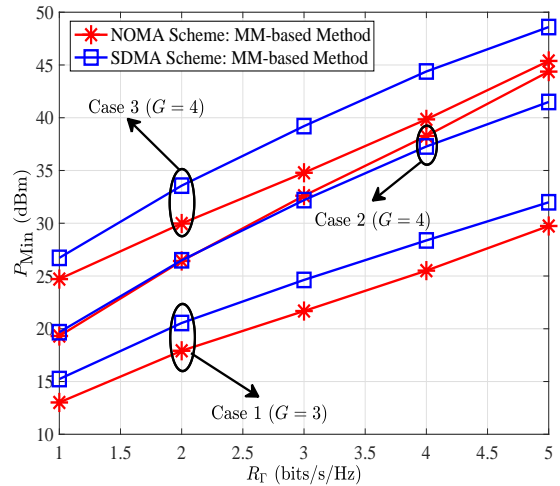}
    %\vspace{-12pt}
    \caption{Average minimum transmit power $P_\textrm{Min}$ (dBm) versus target data rate $R_\Gamma$ (bits/s/Hz) for different numbers of groups $G$. Parameters: $M=10$, $U=3$, $\alpha=2.5$ and $\sigma^2=-80$ dBm. In Case 1, 2 and 3, the region location models shown in Fig. \ref{fig:LocationModel}(A), \ref{fig:LocationModel}(B) and \ref{fig:LocationModel}(C) are used, respectively.}
    \label{fig:PMIN_R_G}
    %\vspace{-12pt}
\end{figure}
For completeness, Fig. \ref{fig:PMIN_R_G} investigates the impact of number of groups $G$ on the performance of NOMA Scheme and SDMA Scheme. Compared with Case 1, an extra user group accesses into the spectrum in Case 2 (or 3), which consequently makes $P_\textrm{Min}$ increase for both transmission schemes. Regarding to the comparison between NOMA Scheme and SDMA Scheme, it is very difficult to determine which scheme is more robust in supporting more groups. This is because the comparison between them depends on the disparity level between any two regions. In particular, if the newly accessing group shares the similar level of large-scale fading with one of the existing groups, it is unfavorable to adopt NOMA Scheme (shown as Case 2). However, if the newly accessing group can still ensure that users coming from any two different regions have remarkable differences in channel gains, the cost of SIC will be relatively small and then the advantage of NOMA can be sufficiently exploited (shown as Case 3).

\subsection{Simulation Results for the Second Scenario}
\begin{figure}[!t]
    \centering
    \hspace{-0.05in}\includegraphics[height = 6.9cm, width = 8.4cm]{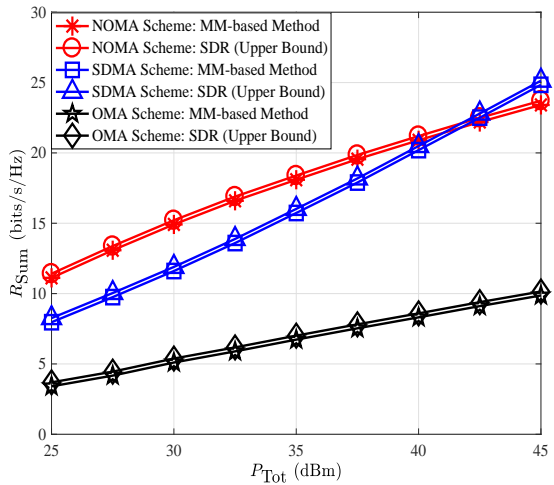}
    %\vspace{-12pt}
    \caption{Average system sum rate $R_\textrm{Sum}$ (bits/s/Hz) versus total transmit power available at the transmitter $P_\textrm{Tot}$. Parameters: $M=8$, $G=3$, $U=3$, $\alpha=2.5$ and $\sigma^2=-80$ dBm.}
    \label{fig:RFair_PtotdBm}
    %\vspace{-12pt}
\end{figure}
Fig. \ref{fig:RFair_PtotdBm} shows average system sum rate $R_\textrm{Sum}$ versus total transmit power available at the transmitter $P_\textrm{Tot}$, under the predefined fair transmission policy, by using the region location model shown in Fig. \ref{fig:LocationModel}(A). It can be seen that when $P_\textrm{Tot}$ is below a certain level, NOMA Scheme has a superior performance compared with SDMA Scheme, but when $P_\textrm{Tot}$ is very large, it is more beneficial to use SDMA Scheme. This is because the transmit data rates of all user groups are increased evenly as $P_\textrm{Tot}$ becomes larger, which makes the cost of SIC increase since the nearer regions need more power to perform SIC, thus restraining the advantage of NOMA. In addition, we can see that NOMA Scheme always has a significant performance improvement in comparison with OMA Scheme, which is due to the reasons similar to those for Fig. \ref{fig:PMIN_R}.

\begin{figure}[!t]
    \centering
    \includegraphics[height = 6.9cm, width = 8.4cm]{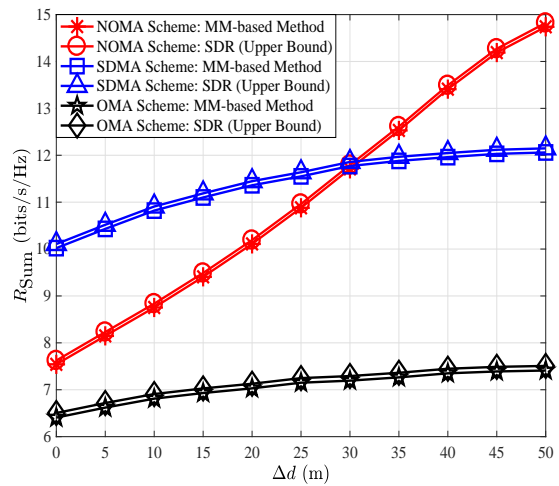}
    %\vspace{-12pt}
    \caption{Average system sum rate $R_\textrm{Sum}$ (bits/s/Hz) versus $\Delta d$ (m), where $\Delta d$ is the difference of the two regions' average distances to the transmitter. Parameters: $P_\textrm{Tot}=30$ dBm, $M=6$, $G=2$, $U=4$, $\alpha=2.5$, $\sigma^2=-80$ dBm.}
    \label{fig:RFair_DeltaDistance}
    %\vspace{-12pt}
\end{figure}
Fig. \ref{fig:RFair_DeltaDistance} investigates the impact of region locations on the system sum rate performance by using the variable region location model shown in Fig. \ref{fig:LocationModel}(D). First of all, as expected, NOMA Scheme is always superior to OMA Scheme. Secondly, we can see that when $\Delta d$ is not large enough, it is not beneficial to apply NOMA Scheme. Further, similar to previous simulation results, NOMA Scheme outperforms SDMA Scheme when $\Delta d$ is sufficiently large, which implies that the natural near-far effect among users greatly benefits NOMA. This implication plausibly verifies that the idiosyncrasy of NOMA is to utilize the SINR imparity among users.

\subsection{Simulation Results for the Third Scenario}
\begin{figure}[!t]
    \centering
    \includegraphics[height = 6.9cm, width = 8.4cm]{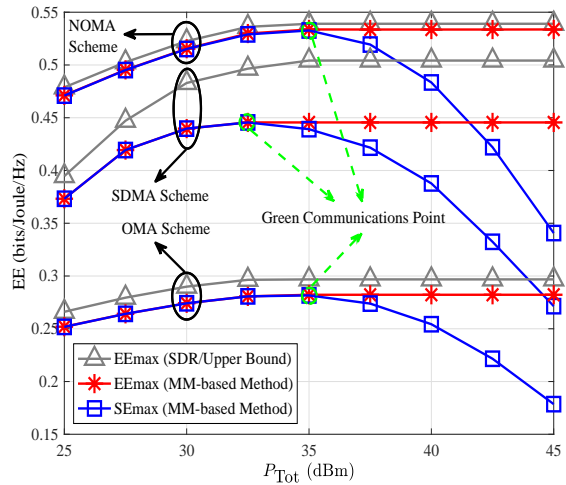}
    %\vspace{-12pt}
    \caption{Average EE (bits/Joule/Hz) versus total transmit power available at the transmitter $P_\textrm{Tot}$. Parameters: $R_\textrm{Pri}=1$ bits/s/Hz, $M=6$, $G=2$, $U=5$, $\alpha=2.5$, $\sigma^2=-80$ dBm and $P_c=44.77$ dBm (30 Watt).}
    \label{fig:CogEE_PtotdBm}
    %\vspace{-12pt}
\end{figure}
Fig. \ref{fig:CogEE_PtotdBm} shows the average EE of the upgraded system under the region location model shown in Fig. \ref{fig:LocationModel}(E). It can be observed that there exists a ``Green Communications Point'' where both ``EEmax'' and ``SEmax'' strategies obtain the maximum EE. When the total power $P_\textrm{Tot}$ is larger than the power of Green Communications Point, using full power is not the optimal strategy from the EE perspective. Besides, NOMA Scheme is definitely more efficient than OMA Scheme in terms of EE. This is because, multiple user groups are simultaneously served with NOMA, which benefits the system in achieving higher diversity gains and more efficient spectrum sharing.

\section{Conclusion}\label{SecCon}
In this work, we have exploited geographical information to both NOMA to and multi-group multicast technologies, which have brought about a novel location-based spectrum sharing framework termed as NOMA assisted multi-region geocast. Specifically, downlink beamforming designs have been investigated for three typical multi-region geocast scenarios in MISO settings. For the accompanying non-convex and intractable problems, efficient algorithms have been proposed based on the MM method. Our comprehensive investigations and simulations have shown that NOMA Scheme is always superior to OMA Scheme in terms of both SE and EE but not necessarily in comparison with SDMA Scheme. In particular, NOMA Scheme achieves better spectrum sharing performance than SDMA Scheme when the disparity level between regions is remarkable. It has been revealed that the nature of NOMA is to exploit channel condition difference among users for enabling multi-user interference cancellation. Furthermore, NOMA Scheme is more favorable for limited transmit antennas and massive connectivity. These all show that intelligent usages of NOMA assisted multi-region geocast can bring tremendous promise for enhancing spectrum sharing in future 5G wireless systems.

\end{document}